\documentclass[twocolumn]{aastex631}

\usepackage{graphicx}	
\usepackage{amsmath}	
\usepackage{subfigure}
\usepackage{natbib}
\usepackage{color}
\usepackage{tabularx}
\usepackage{longtable}
\usepackage{bm}
\usepackage{threeparttable}
\usepackage{enumerate}
\usepackage[normalem]{ulem}
\usepackage{verbatim}

\newcommand{\code}[1]{{\texttt{#1}}}
\newcommand{\tess}{{\it TESS}}
\newcommand{\kepler}{{\it Kepler}}

\newcommand{\gaia}{{\it Gaia}}

\newcommand{\tar}{{TOI-4201}}

\begin{document}

\title{A massive hot Jupiter orbiting a metal-rich early-M star discovered in the TESS full frame images}

\correspondingauthor{Tianjun Gan}
\email{tianjungan@gmail.com}

\author[0000-0002-4503-9705]{Tianjun~Gan}
\affil{Department of Astronomy, Tsinghua University, Beijing 100084, People's Republic of China}

\author[0000-0001-9291-5555]{Charles Cadieux}
\affiliation{Universit\'e de Montr\'eal, D\'epartement de Physique, IREX, Montr\'eal, QC H3C 3J7, Canada}

\author[0000-0003-0029-2835]{Farbod Jahandar} 
\affiliation{Universit\'e de Montr\'eal, D\'epartement de Physique, IREX, Montr\'eal, QC H3C 3J7, Canada}

\author[0000-0001-9504-3174]{Allona Vazan} 
\affil{Astrophysics Research Center (ARCO), Dept. of Natural Sciences, Open University of Israel, Raanana, 4353701 Israel}

\author[0000-0002-6937-9034]{Sharon X. Wang}
\affil{Department of Astronomy, Tsinghua University, Beijing 100084, People's Republic of China}

\author[0000-0001-8317-2788]{Shude Mao}
\affil{Department of Astronomy, Tsinghua University, Beijing 100084, People's Republic of China}
\affil{National Astronomical Observatories, Chinese Academy of Sciences, 20A Datun Road, Chaoyang District, Beijing 100012, People's Republic of China}

\author[0000-0003-0353-9741]{Jaime A. Alvarado-Montes} 
\affil{School of Mathematical and Physical Sciences, Macquarie University, Balaclava Road, North Ryde, NSW 2109, Australia}
\affil{The Macquarie University Astrophysics and Space Technologies Research Centre, Macquarie University, Balaclava Road, North Ryde, NSW 2109, Australia}

\author[0000-0001-5466-4628]{D. N. C. Lin}
\affil{Department of Astronomy and Astrophysics University of California, Santa Cruz, CA 95064, USA}
\affil{Institute for Advanced Studies Tsinghua University, Beijing 100086, People’s Republic of China}

\author[0000-0003-3506-5667]{\'Etienne Artigau}
\affiliation{Universit\'e de Montr\'eal, D\'epartement de Physique, IREX, Montr\'eal, QC H3C 3J7, Canada}
\affiliation{Observatoire du Mont-M\'egantic, Universit\'e de Montr\'eal, Montr\'eal, QC H3C 3J7, Canada}
\author[0000-0003-4166-4121]{Neil J. Cook}
\affiliation{Universit\'e de Montr\'eal, D\'epartement de Physique, IREX, Montr\'eal, QC H3C 3J7, Canada}
\author[0000-0001-5485-4675]{Ren\'e Doyon}
\affiliation{Universit\'e de Montr\'eal, D\'epartement de Physique, IREX, Montr\'eal, QC H3C 3J7, Canada}
\affiliation{Observatoire du Mont-M\'egantic, Universit\'e de Montr\'eal, Montr\'eal, QC H3C 3J7, Canada}

\author[0000-0003-3654-1602]{Andrew W. Mann} 
\affiliation{Department of Physics and Astronomy, The University of North Carolina at Chapel Hill, Chapel Hill, NC 27599-3255, USA}

\author[0000-0002-3481-9052]{Keivan G. Stassun} 
\affil{Department of Physics and Astronomy, Vanderbilt University, 6301 Stevenson Center Ln., Nashville, TN 37235, USA}
\affil{Department of Physics, Fisk University, 1000 17th Avenue North, Nashville, TN 37208, USA}

\author[0000-0002-6523-9536]{Adam J.\ Burgasser}
\affiliation{Department of Astronomy \& Astrophysics, UC San Diego, 9500 Gilman Drive, La Jolla, CA 92093, USA}

\author[0000-0002-3627-1676]{Benjamin V.\ Rackham}
\altaffiliation{51 Pegasi b Fellow}
\affil{Department of Earth, Atmospheric and Planetary Science, Massachusetts Institute of Technology, 77 Massachusetts Avenue, Cambridge, MA 02139, USA}
\affil{Kavli Institute for Astrophysics and Space Research, Massachusetts Institute of Technology, 77 Massachusetts Avenue, Cambridge, MA 02139, USA}

\author[0000-0002-2532-2853]{Steve~B.~Howell} 
\affiliation{NASA Ames Research Center, Moffett Field, CA 94035, USA}

\author[0000-0001-6588-9574]{Karen A. Collins} 
\affil{Center for Astrophysics ${\rm \mid}$ Harvard {\rm \&} Smithsonian, 60 Garden Street, Cambridge, MA 02138, USA}

\author[0000-0003-1464-9276]{Khalid Barkaoui} 
\affil{Astrobiology Research Unit, Universit\'e de Li\`ege, 19C All\'ee du 6 Ao\^ut, 4000 Li\`ege, Belgium}
\affil{Department of Earth, Atmospheric and Planetary Science, Massachusetts Institute of Technology, 77 Massachusetts Avenue, Cambridge, MA 02139, USA}
\affil{Instituto de Astrof\'isica de Canarias (IAC), Calle V\'ia L\'actea s/n, 38200, La Laguna, Tenerife, Spain}

\author[0000-0002-1836-3120]{Avi Shporer} 
\affil{Department of Physics and Kavli Institute for Astrophysics and Space Research, Massachusetts Institute of Technology, Cambridge, MA 02139, USA}

\author[0000-0002-6424-3410]{Jerome de Leon}
\affiliation{Department of Multi-Disciplinary Sciences, Graduate School of Arts and Sciences, The University of Tokyo, 3-8-1 Komaba, Meguro, Tokyo
153-8902, Japan}

\author{Luc Arnold}
\affiliation{Canada–France–Hawaii Telescope, 65-1238 Mamalahoa Hwy, Kamuela, HI 96743, USA}

\author{George~R.~Ricker}
\affil{Department of Physics and Kavli Institute for Astrophysics and Space Research, Massachusetts Institute of Technology, Cambridge, MA 02139, USA}

\author{Roland~Vanderspek}
\affil{Department of Physics and Kavli Institute for Astrophysics and Space Research, Massachusetts Institute of Technology, Cambridge, MA 02139, USA}

\author{David~W.~Latham}
\affil{Center for Astrophysics ${\rm \mid}$ Harvard {\rm \&} Smithsonian, 60 Garden Street, Cambridge, MA 02138, USA}

\author[0000-0002-6892-6948]{Sara~Seager}
\affil{Department of Physics and Kavli Institute for Astrophysics and Space Research, Massachusetts Institute of Technology, Cambridge, MA 02139, USA}
\affil{Department of Earth, Atmospheric and Planetary Science, Massachusetts Institute of Technology, 77 Massachusetts Avenue, Cambridge, MA 02139, USA}
\affil{Department of Aeronautics and Astronautics, MIT, 77 Massachusetts Avenue, Cambridge, MA 02139, USA}

\author{Joshua~N.~Winn}
\affil{Department of Astrophysical Sciences, Princeton University, 4 Ivy Lane, Princeton, NJ 08544, USA}

\author{Jon~M.~Jenkins}
\affil{NASA Ames Research Center, Moffett Field, CA 94035, USA}


\author[0000-0001-9892-2406]{Artem Burdanov} 
\affil{Department of Earth, Atmospheric and Planetary Science, Massachusetts Institute of Technology, 77 Massachusetts Avenue, Cambridge, MA 02139, USA}

\author[0000-0002-9003-484X]{David~Charbonneau}
\affil{Center for Astrophysics ${\rm \mid}$ Harvard {\rm \&} Smithsonian, 60 Garden Street, Cambridge, MA 02138, USA}

\author{Georgina Dransfield} 
\affil{School of Physics \& Astronomy, University of Birmingham, Edgbaston, Birmingham B15 2TT, UK}

\author[0000-0002-4909-5763]{Akihiko Fukui}
\affiliation{Komaba Institute for Science, The University of Tokyo, 3-8-1 Komaba, Meguro, Tokyo 153-8902, Japan}
\affiliation{Instituto de Astrof\'{i}sica de Canarias (IAC), 38205 La Laguna, Tenerife, Spain}

\author[0000-0001-9800-6248]{Elise Furlan}
\affiliation{NASA Exoplanet Science Institute, Caltech/IPAC, Mail Code 100-22, 1200 E. California Blvd., Pasadena, CA 91125, USA}

\author[0000-0003-1462-7739]{Micha{\" e}l Gillon}
\affil{Astrobiology Research Unit, Universit\'e de Li\`ege, 19C All\'ee du 6 Ao\^ut, 4000 Li\`ege, Belgium}

\author[0000-0003-0030-332X]{Matthew J. Hooton} 
\affiliation{Cavendish Laboratory, JJ Thomson Avenue, Cambridge CB3 0HE, UK}


\author[0000-0002-7871-085X]{Hannah M.~Lewis}
\affiliation{Space Telescope Science Institute, 3700 San Martin Drive, Baltimore, MD, 21218, USA}

\author[0000-0001-7746-5795]{Colin Littlefield}
\affiliation{Bay Area Environmental Research Institute, Moffett Field, CA 94035, USA}
\affiliation{NASA Ames Research Center, Moffett Field, CA 94035, USA}

\author[0000-0002-4510-2268]{Ismael~Mireles}
\affiliation{Department of Physics and Astronomy, University of New Mexico, 210 Yale Blvd NE, Albuquerque, NM 87106, USA}

\author[0000-0001-8511-2981]{Norio Narita}
\affiliation{Komaba Institute for Science, The University of Tokyo, 3-8-1 Komaba, Meguro, Tokyo 153-8902, Japan}
\affiliation{Astrobiology Center, 2-21-1 Osawa, Mitaka, Tokyo 181-8588, Japan}
\affiliation{Instituto de Astrof\'{i}sica de Canarias (IAC), 38205 La Laguna, Tenerife, Spain}

\author[0000-0003-4672-8411]{Chris W. Ormel}
\affil{Department of Astronomy, Tsinghua University, Beijing 100084, People's Republic of China}

\author[0000-0002-8964-8377]{Samuel~N.~Quinn}
\affil{Center for Astrophysics ${\rm \mid}$ Harvard {\rm \&} Smithsonian, 60 Garden Street, Cambridge, MA 02138, USA}

\author[0000-0003-3904-6754]{Ramotholo Sefako} 
\affiliation{South African Astronomical Observatory, P.O. Box 9, Observatory, Cape Town 7935, South Africa}

\author{Mathilde Timmermans} 
\affil{Astrobiology Research Unit, Universit\'e de Li\`ege, 19C All\'ee du 6 Ao\^ut, 4000 Li\`ege, Belgium}

\author{Michael~Vezie}
\affiliation{Department of Physics and Kavli Institute for Astrophysics and Space Research, Massachusetts Institute of Technology, Cambridge, MA 02139, USA}

\author[0000-0003-2415-2191]{Julien de Wit}
\affil{Department of Earth, Atmospheric and Planetary Science, Massachusetts Institute of Technology, 77 Massachusetts Avenue, Cambridge, MA 02139, USA}




\begin{abstract}

Observations and statistical studies have shown that giant planets are rare around M dwarfs compared with Sun-like stars. The formation mechanism of these extreme systems remains under debate for decades. With the help of the TESS mission and ground based follow-up observations, we report the discovery of \tar b, the most massive and densest hot Jupiter around an M dwarf known so far with a radius of $1.22\pm 0.04\ R_J$ and a mass of $2.48\pm0.09\ M_J$, about 5 times heavier than most other giant planets around M dwarfs. It also has the highest planet-to-star mass ratio ($q\sim 4\times 10^{-3}$) among such systems. The host star is an early-M dwarf with a mass of $0.61\pm0.02\ M_{\odot}$ and a radius of $0.63\pm0.02\ R_{\odot}$. It has significant super-solar iron abundance ([Fe/H]=$0.52\pm 0.08$ dex). However, interior structure modeling suggests that its planet \tar b is metal-poor, which challenges the classical core-accretion correlation of stellar-planet metallicity, unless the planet is inflated by additional energy sources. Building on the detection of this planet, we compare the stellar metallicity distribution of four planetary groups: hot/warm Jupiters around G/M dwarfs. We find that hot/warm Jupiters show a similar metallicity dependence around G-type stars. For M dwarf host stars, the occurrence of hot Jupiters shows a much stronger correlation with iron abundance, while warm Jupiters display a weaker preference, indicating possible different formation histories.


\end{abstract}



\section{Introduction} \label{sec:intro}

As the most abundant stellar population in our Milky Way \citep{Henry2006}, M dwarfs are popular targets for exoplanet research. The \kepler\ mission \citep{Borucki2010} has revealed that early-M stars host close-in small planets about 3 times more frequent than Sun-like stars \citep{Petigura2013,Dressing2013,Dressing2015,Gaidos2016}. At the same time, however, recent work from \cite{Gan2023} found that hot Jupiters (defined as orbital period $P\leq 10$ days, radius $R_{p}\geq 7\ R_{\oplus}$) are depleted around early-M dwarfs with an occurrence rate of $0.27\pm0.09\%$ \citep[see also][]{Bryant2023}, in contrast to the frequency $\sim 0.6\%$ around FGK stars \citep[e.g.,][]{Fressin2013,Petigura2018,Zhou2019,Beleznay2022}. This difference is even more significant if comparing with the result $\sim 1.0$\% around FGK stars from radial velocity (RV) surveys \citep[e.g.,][]{Mayor2011,Cumming2008,Wright2012,Wittenmyer2020}.

The rarity of giant planets around M dwarfs is probably a natural outcome of the core accretion planet formation mechanism \citep{Pollack1996}. Under this hypothesis, the low mass of the protoplanetary disk around a low-mass star makes it difficult to form a massive enough solid core to start runaway gas accretion before disk dissipation \citep{Laughlin2004,Ida2005,Kennedy2008}. Indeed, several simulation works based on this paradigm reported a similar increasing trend of hot Jupiter occurrence rate as the stellar mass going up from 0.5 towards 1 $M_{\odot}$ \citep[e.g.,][]{Liu2019,Burn2021}. In addition, some initial studies found that most confirmed hot Jupiters are orbiting metal-rich M dwarfs with a median metallicity around 0.3 dex, which is higher than that of Sun-like stars harboring gas giants \citep{Gantoi530,Kanodia2022,Kagetani2023}. This is also supposed to be evidence in line with the core accretion scenario as the stellar metallicity is generally correlated with the mass of solids in the protoplanetary disk that are available for forming planets \citep{Santos2004,Fischer2005,Sousa2011,WWang2018}. In turn, the core accretion framework probably simultaneously explains the fact that small planets are common around M dwarfs because a forming outer giant planet around an FGK star may suppress the formation of close-in super-Earths due to the cut-off of the inward flow of solids \citep{Mulders2021}. 

On the other hand, there are still two relevant types of planetary systems that remain challenging for core accretion. The first population is gas giants around mid-to-late M dwarfs which theoretical works do not expect to form directly \citep[e.g.,][]{Liu2019,Burn2021}, unless a massive core was first formed through, for example, planet-planet collisions \citep{Frelikh2019}. Nevertheless, a few such systems were recently discovered \citep{Morales2019,Hobson2023,Kanodia2023,Kagetani2023}. Additionally, \cite{Bryant2023} also reported an occurrence rate of about 0.1\% for hot Jupiters orbiting stars with $M_\ast \leq 0.4\ M_{\odot}$, indicating that they are not vanishingly rare. Another group is cold Jupiters around M dwarfs. In particular, gas giants beyond the ice line are frequently discovered through the microlensing method \citep{Mao1991,Suzuki2016}. By comparing the observations and predictions from planet population synthesis models, \cite{Schlecker2022} suggested that core accretion theories cannot reproduce the RV-detected giant planets around M dwarfs. Moreover, cold gas giants around M dwarfs detected by RV surveys show a much weaker metallicity dependence compared with hot Jupiters \citep{Gantoi530}, which hints that they follow a different formation path. In these cases, gravitational instability probably plays a role \citep{Boss2000}. 

The \tess\ \citep[Transiting Exoplanet Survey Satellite;][]{Ricker2015} mission, which performs a full-sky photometric survey, has been rapidly increasing the number of gas giants around M dwarfs. Plenty of hot Jupiters have been detected over the last three years (e.g., HATS-71b, \citealt{Bakos2020}; HATS-74Ab and HATS-75b, \citealt{Jordan2022}; TOI-3629b and TOI-3714b, \citealt{Canas2022}). They have enlarged the sample size of short-period giant planets around M stars by a factor of three. A few warm Jupiters were also reported \citep[e.g.,][]{Canas2020}. All of these findings gradually enable glimpses of their formation and evolution history. 

In this manuscript, we report the discovery and confirmation of \tar b, a massive and dense hot Jupiter around an early-M star, which was previously classified as a verified planet candidate by \cite{Gan2023}. The rest of the paper is organized as follows. In Section~\ref{sec:obs}, we detail all observations we use to confirm the planetary nature of \tar b. We summarize the stellar properties in Section~\ref{sec:stellar}. Section~\ref{sec:fit} describes the joint-fit analysis to derive the planet parameters. In Section~\ref{sec:dis}, we discuss the interior structure of \tar b, tidal evolution analysis, the prospects for future characterizations as well as the comparison of stellar metallicity distribution of hot/warm Jupiters around G/M stars. We conclude with our findings in Section~\ref{sec:con}.

\section{Observations} \label{sec:obs}
\subsection{TESS}
\tar\ (TIC 95057860) was first monitored by TESS in Sector 6 between 2018 December 15 and 2019 January 6 with the 30-minute cadence mode, and it was revisited every 10 minutes in Sector 33 from 2020 December 18 to 2021 January 13 during the first extended mission. The full frame images (FFIs) of \tar\ from Sector 6 were processed by the Quick Look Pipeline \citep[QLP;][]{QLP2020a,QLP2020b,Fausnaugh2020}. Due to the faintness of \tar\ ($T{\rm mag}=13.5$), the light curve was not examined initially because the TESS vetting team only inspects targets with $T{\rm mag}\leq 10.5$. Its transiting signal was alerted by the QLP faint-star search program \citep{Kunimoto2022}, which searches for planet candidates around dimmer stars with $10.5\leq T{\rm mag}\leq 13.5$ that have QLP light curves. 

\begin{figure*}
\centering
\subfigure[Full frame image from Sector 6]{\includegraphics[width=0.49\textwidth]{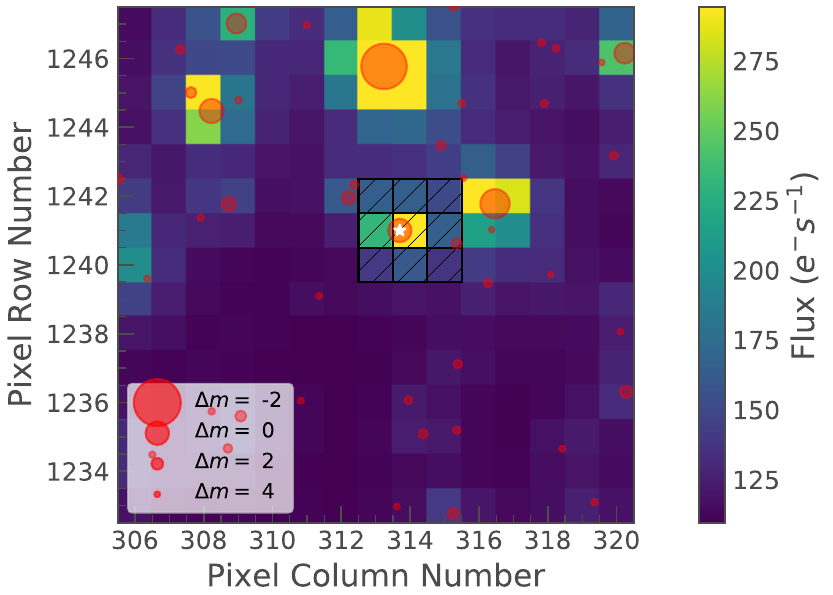}}
\subfigure[Full frame image from Sector 33]{\includegraphics[width=0.49\textwidth]{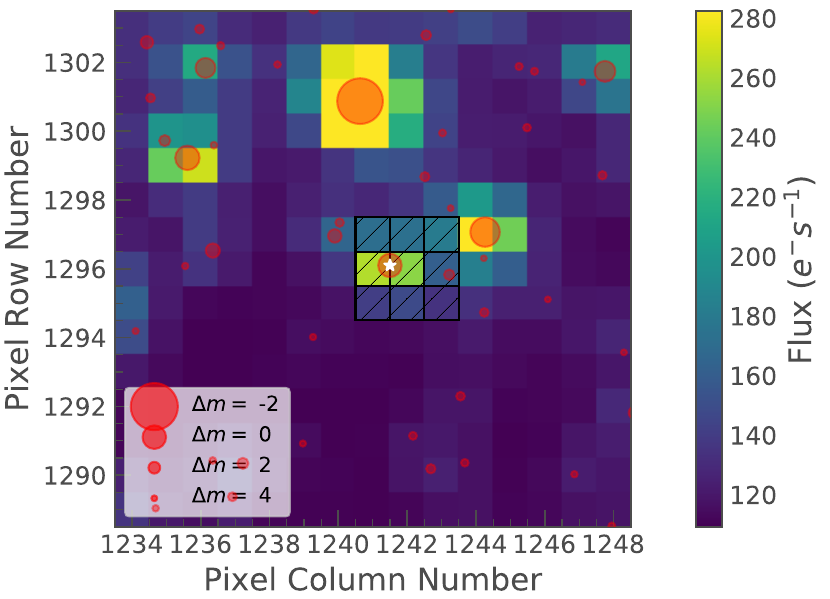}}
\subfigure[TESS light curve from Sector 6]{\includegraphics[width=0.49\textwidth]{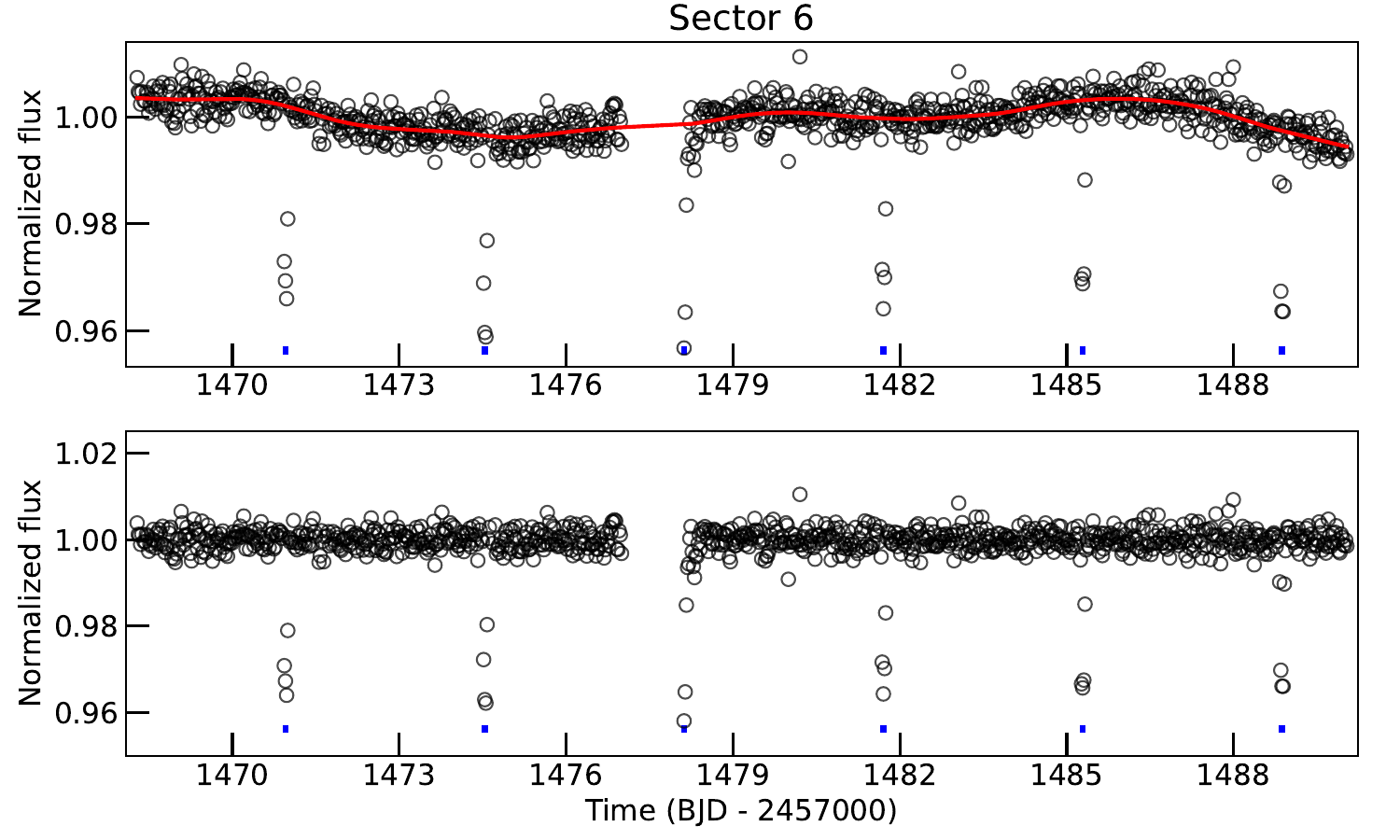}}
\subfigure[TESS light curve from Sector 33]{\includegraphics[width=0.49\textwidth]{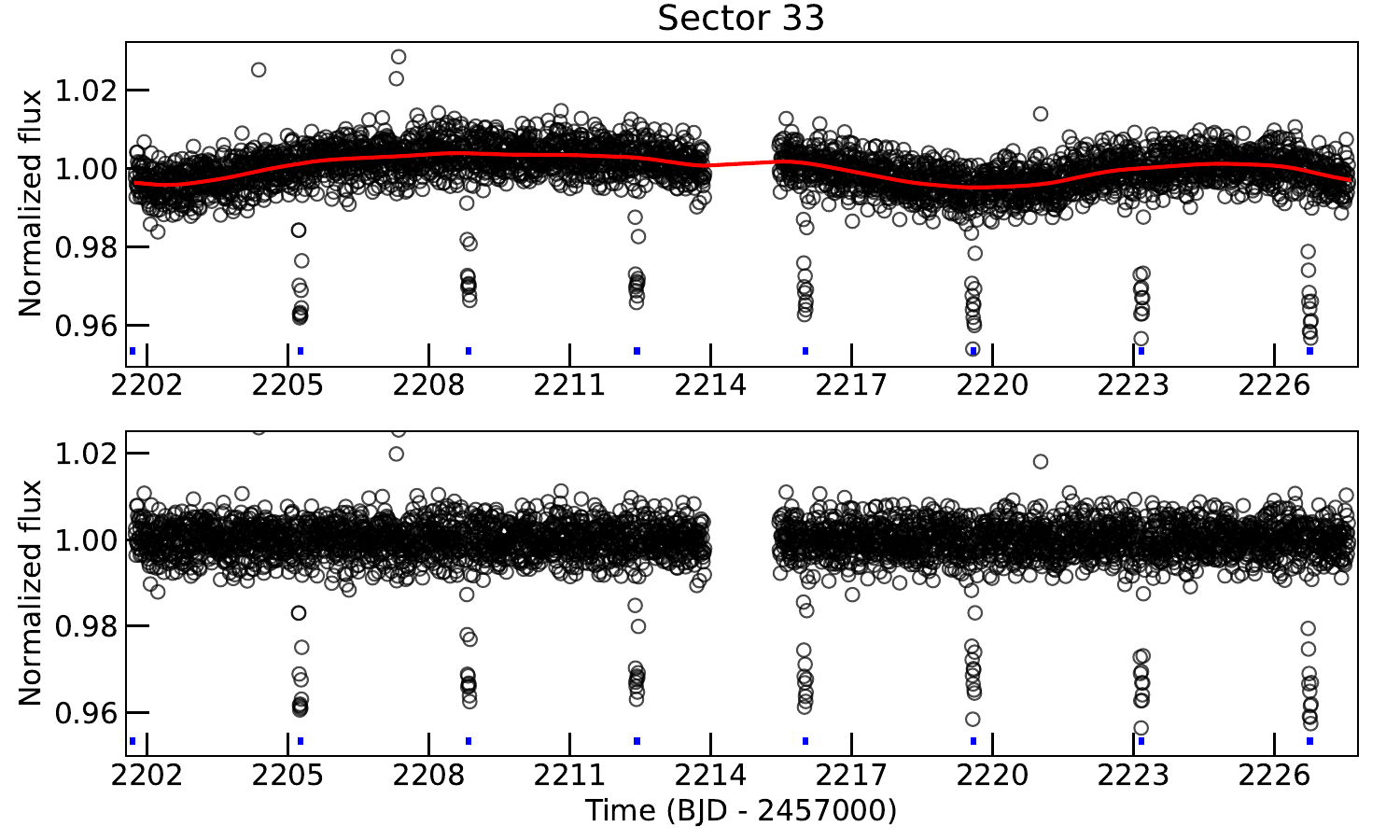}}
\caption{{\it Top panels:} The 15x15 pixel TESS full frame image from Sectors 6 and 33. The black-shaded region represents the 3x3 pixel custom aperture we used to conduct photometry. \tar\ is marked as a white star at the center. Different sizes of red circles represent different magnitudes in contrast with \tar. {\it Bottom panels:} The normalized TESS light curves of \tar\ from Sectors 6 (left) and 33 (right) that we extracted. The red solid lines are the GP models used for detrending. The detrended light curves are shown below. The transits of \tar b are marked in blue ticks.} 
\label{FOV_aper_TESS}
\end{figure*}

However, we note that the light curve of \tar\ from Sector 33 was not produced by QLP. Therefore, we carried out an independent and uniform aperture photometry for both Sectors using the \code{lightkurve} package \citep{lightkurvecollaboration,lightkurve}. We first downloaded the $15\times15$ pixels FFI cutouts of \tar\ from each Sector (see the top panels of Figure \ref{FOV_aper_TESS}), and extracted a raw target light curve using a $3\times3$ pixel custom aperture, centering at the location of \tar. Next, we excluded a $5\times5$ pixel region around the target and constructed a $0.001\sigma$ aperture mask to estimate the background flux. In this way, we picked up all pixels with flux smaller than 0.001 times the standard deviation of flux within the overall $15\times15$-pixel region. Finally, we corrected the sky background variation by subtracting the background flux from the raw target flux. We show the light curves we obtained in Figure \ref{FOV_aper_TESS}.

Combining all TESS data, we performed a transit search using the Transit Least Squares (TLS; \citealt{Hippke2019}) algorithm after smoothing the light curve using a median filter with a window size of 0.3 days. We did not find additional significant periodic signals other than the 3.58-day one from \tar b. After masking out all in-transit data, we fit a Gaussian Process (GP) model with a Mat\'{e}rn-3/2 kernel using \code{celerite} \citep{Foreman2017} for the light curve from each Sector. We divided the light curve by the best-fit GP model for detrending and normalization. 

\subsection{Ground-based photometry}

In order to rule out false positive scenarios such as nearby and blended eclipsing binary, and refine the transit ephemeris as well as transit depth measurement, we collected a total of eight ground-based follow-up light curves for \tar, as part of the TESS Follow-up Observing Program (TFOP). We scheduled all observations based on the transit information from the \tess\ Transit Finder (TTF), which is a customized version of the \code{Tapir} software package \citep{Jensen2013}. We summarize the details of all observations in Table~\ref{po}, and describe each of them below. 

\subsubsection{LCOGT}

We acquired two SDSS $g'$ and $i'$ band alternating time-series observations using 1-meter telescopes from the Las Cumbres Observatory Global Telescope network \citep[LCOGT;][]{Brown2013}. The first partial transit (ingress-only) observation was done at South African Astronomical Observatory (SAAO) on 2021 September 26, while the second one was obtained at Cerro Tololo Interamerican Observatory (CTIO) on 2021 October 4 covering the full event. Both observations were carried out with the Sinistro cameras, which have a field of view of $26'\times26'$ and a plate scale of $0.389\arcsec$/pixel. The exposure times of $g'$ and $i'$ band observations are 300s and 180s, respectively. After the raw images were calibrated by the automatic \code{BANZAI} pipeline \citep{McCully2018}, we conducted a photometric analysis using the \code{AstroImageJ} software \citep{Collins2017} with a 22-pixel ($8.5\arcsec$) and 16-pixel ($6.2\arcsec$) aperture for two observations. We confirmed the transit signal on target at the ephemeris provided by TESS. 

\subsubsection{MuSCAT}

We observed a full transit of \tar b on 2022 Jan 30 with MuSCAT using exposure times of 120, 50, 50~s for $g$,  $r$, and  $z_s$ bands, respectively. MuSCAT is a multi-band simultaneous camera installed on the 188-cm telescope of the National Astronomical Observatory of Japan (NAOJ) in Okayama, Japan \citep{Narita2015}. It has three 1k CCDs each with $6.1' \times 6.1'$ FOV, enabling simultaneous photometry in the $g$ (400–550 nm), $r$ (550–700 nm), and $z_s$ (820–920 nm) bands. The data reduction and differential photometry were performed using the pipeline described in \citet{Fukui2011}. We optimized both the aperture radii and the set of comparison stars by minimizing the dispersion of the resulting relative light curves. Five comparison stars and aperture radii of 12 pixels (4.2$\arcsec$) yield the optimum light curves for all bands. 

\subsubsection{SPECULOOS-North}

\begin{figure}
\includegraphics[width=0.49\textwidth]{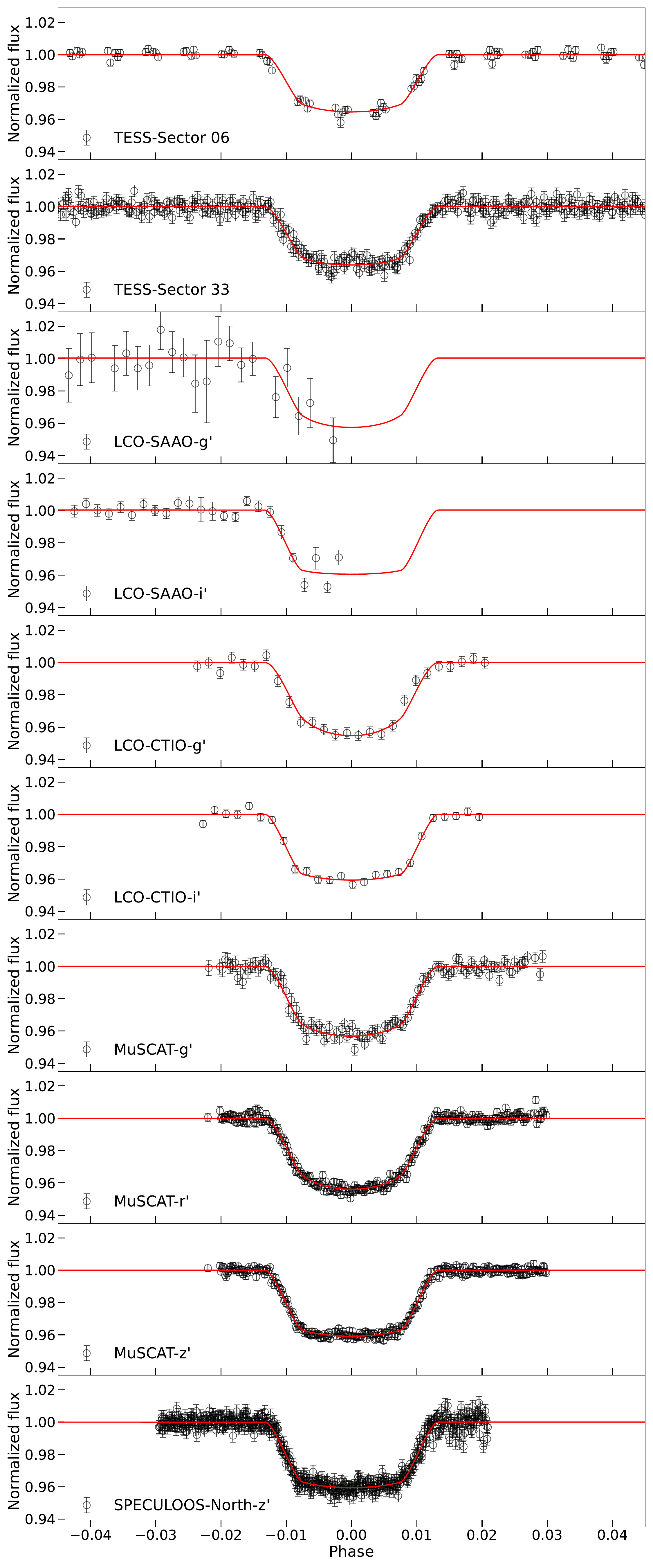}
\caption{The TESS and ground-based light curves folded in phase with the orbital period of \tar b. The red solid lines are the best-fit transit models. Instrument name and filter information are given at the bottom left of each panel.}
\label{ground}
\end{figure}

We obtained a full transit of TOI-4201\,b by the SPECULOOS-North telescope on 2023 February 4 in the Sloan-$z'$ filter with an exposure time of 20s. SPECULOOS-North is a 1.0-m Ritchey-Chretien telescope equipped with a thermoelectrically cooled 2K$\times$2K Andor iKon-L BEX2-DD CCD camera with a pixel scale of $0.35\arcsec$ and a field-of-view of $12\arcmin\times12\arcmin$ \citep{Burdanov2022}. It is a twin of the SPECULOOS-South located in ESO-Paranal in Chile \citep{Jehin2018Msngr,Delrez2018,Sebastian_2021AA} and SAINT-EX located at the Sierra de San Pedro M\'artir in Baja California, M\'exico \citep{Demory_AA_SAINTEX_2020}. Data reduction and  photometric measurements were performed using the {\tt PROSE}\footnote{\url{https://github.com/lgrcia/prose}} pipeline \citep{garcia2021}. The resulting light curve is presented in Figure~\ref{ground}.

\begin{table*}
    \centering
    \caption{Ground-based photometric follow-up observations for \tar}
    \begin{tabular}{ccccccc}
        \hline\hline
        Obs Date &Telescope &Filter &Pixel Scale (arcsec) & FWHM (arcsec) &\# of observations &Coverage\\ \hline
        2021-09-26 &LCO-SAAO &$g'$ \& $i'$ &0.39 &5.1 &23 \& 24 &Ingress\\
        2021-10-04 &LCO-CTIO &$g'$ \& $i'$ &0.39 &3.4 &26 \& 25 &Full\\
        2022-01-30 &MuSCAT &$g'$ \& $r'$ \& $z'$ &0.36 &3.0 &124 \& 301 \& 299 &Full\\
        2023-02-04 &SPECULOOS-North &$z'$ &0.35 &1.0 &519 &Full\\
         \hline
    \end{tabular}
    \label{po}
\end{table*}

\subsection{Spectroscopic follow-up}

\subsubsection{MagE}

We obtained medium-resolution optical spectra of \tar\ on 2022 October 06 (UT) with the Magellan Echellette (MagE) spectrograph \citep{2008SPIE.7014E.169M} on the 6.5-m Magellan Baade Telescope.
Conditions were clear with seeing of $0\farcs7$, and we observed the target during nautical twilight.
We used the 0$\farcs$7$\times$10$\arcsec$ slit to obtain resolution $\lambda/\Delta\lambda$ $\approx$ 6000 over 3200--10\,000\,{\AA}.
We collected two, 290-s exposures at an airmass of 1.04.
We also observed the spectrophotometric calibrator Feige 110 during the night for flux calibration \citep{1992PASP..104..533H,1994PASP..106..566H}, and obtained bias exposures as well as ThAr arc lamp and Xe flash and incandescent flatfield lamp exposures at the start of the night for wavelength and flux calibration, respectively. 
We did not observe a telluric absorption calibrator for these observations, hence telluric features remain in the spectra.
Data were reduced using PypeIt \citep{pypeit:joss_pub,pypeit:zenodo} using standard settings. The final calibrated spectrum has a median signal-to-noise $\approx$100 in the 8000\,{\AA} region. 

\begin{figure}
\includegraphics[width=0.49\textwidth]{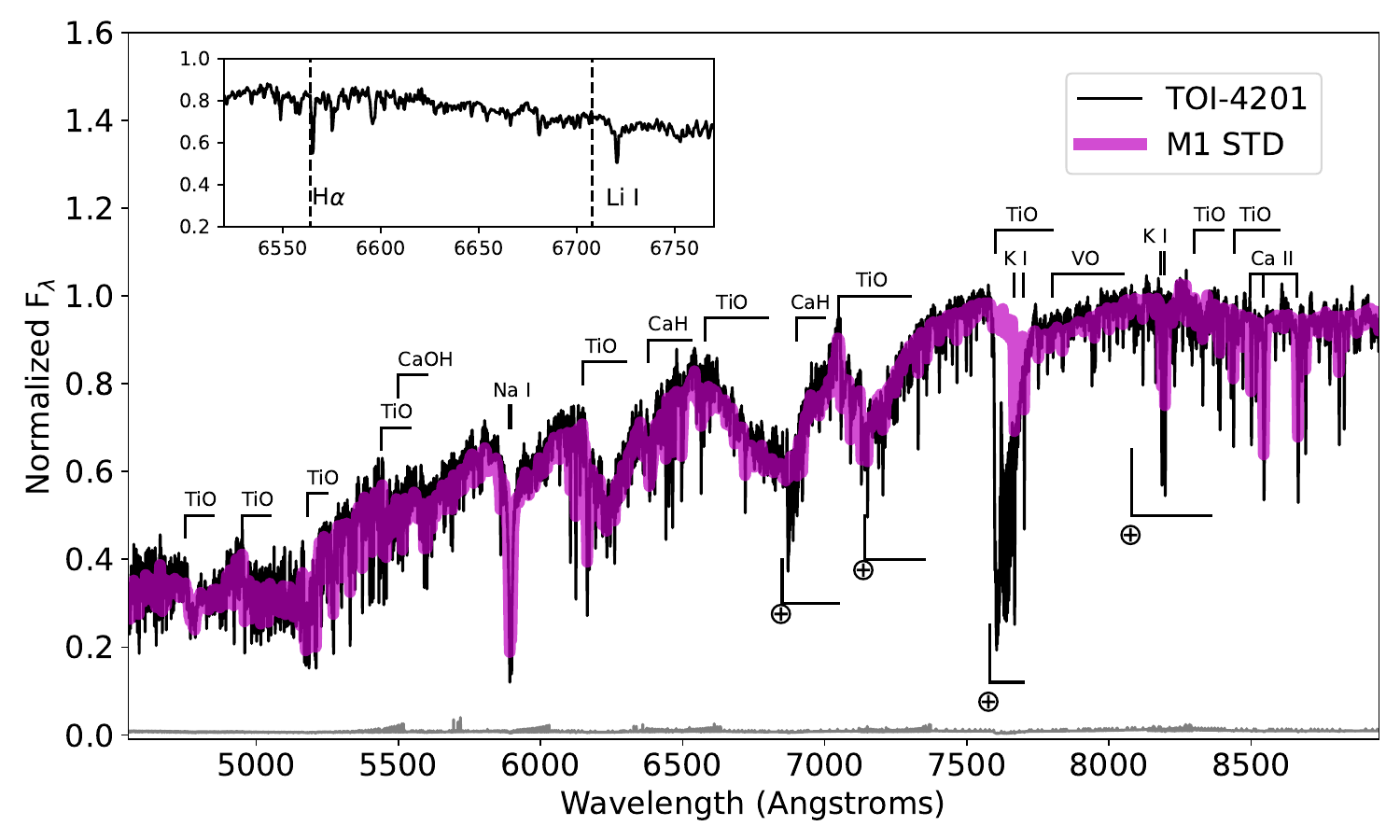}
\caption{MagE optical spectrum of \tar\ in normalized $F_\lambda$ units (black line, uncertainties in grey), compared to the best-match M1 dwarf spectral template (magenta line; data from \citealt{2007AJ....133..531B}). We also label typical spectral absorption features for M dwarfs and regions of uncorrected telluric absorption ($\oplus$). The inset box shows a close-up of the 6530--6780~{\AA} region indicating H$\alpha$ in absorption and no Li~I detection.}
\label{fig:mage}
\end{figure}

Figure~\ref{fig:mage} compares the MagE spectrum of \tar\ to the best-match spectral template constructed from Sloan Digital Sky Survey data \citep{2007AJ....133..531B}. Note that the template has a lower mean resolution ($\lambda/\Delta\lambda$ $\approx$ 2000), but is nonetheless an excellent match, indicating a dwarf classification. This template match is confirmed by index-based classifications from \citet{1995AJ....110.1838R,1997AJ....113..806G}; and \citet{2003AJ....125.1598L}, which span M0.5--M1.5. 
We therefore adopt a spectral classification of M1.0$\pm$0.5 for this source.
Inspection of the data show H$\alpha$ in absorption with an equivalent width of 0.426$\pm$0.017~{\AA}, and no detection of Li~I, ruling out a young ($\lesssim$10~Myr) source.
Using the calibration of \citet{2013AJ....145..102L}, we measure a $\zeta$ metallicity index of 1.166$\pm$0.008, indicating a slightly supersolar metallicity. The metallicity relation of \citet{2013AJ....145...52M} yields [Fe/H] = 0.22$\pm$0.20 for this $\zeta$ value, where the uncertainty is dominated by systematics among the calibration sample. We conduct a more detailed abundance analysis using the SPIRou data in the following.

\subsubsection{SPIRou}

We obtained a total of 28 spectroscopic observations for \tar\ using SPIRou \citep[SpectroPolarim\`etre InfraROUge;][]{Donati2020} between 2022 November 6 and 2023 January 1. SPIRou is a near-infrared high resolution ($R\approx75,000$) fiber-fed echelle spectrograph installed on the 3.6m Canada-France-Hawaii Telescope (CFHT), covering a wavelength range from 0.98 to 2.5\,$\mu$m. Since \tar\ is a faint target ($H{\rm mag} > 11$), instead of taking simultaneous drift calibration with the thermalized Farby-P\'erot (FP) etalon, we opted to use the Dark mode to avoid contaminating science spectra. We conducted all of our observations under an environment of airmass around 1.3 and seeing about $0.6\arcsec$. Every night, we took two continuous sequences of \tar\ at a random phase, with a 1200s exposure time for each one. For our 28 spectra, the median signal-to-noise ratio (SNR) per pixel in the middle of $H$ band is about 40.

The SPIRou data were reduced using \code{APERO} version 0.7.275 \citep{Cook2022}. The main steps of data reduction are described in the following but readers are referred to \cite{Cook2022} for a complete description of the \code{APERO} pipeline and modules. Raw frames (science and calibration) are first pre-processed to remove known structures on the 4096$\times$4096 detector. Nightly calibration images are taken to locate the echelle orders and to perform a series of corrections (e.g., flat, thermal background, blaze, etc.). The flux is extracted from science and calibration frames to produce per-order spectra. A nightly pixel-to-wavelength solution is derived following the method of \cite{Hobson2021}. Finally, a three steps telluric correction is applied to the science data. A first cleaning is done with a TAPAS \citep{Bertaux2014} atmospheric transmission model. The remaining residuals (percent-level in deep H$_2$O bands) are then fitted in this second step using a grid of telluric models generated from observations of hot stars (fast rotators) with SPIRou covering ranges of airmass and water column. The last step mitigates finite resolution effects, i.e., the fact that deconvolution between stellar+telluric spectrum with instrumental profile is always imperfect \citep{Wang2022telluric}. This telluric correction method yields final residuals at the level of the PCA-based approach of \cite{Artigau2014} and will be described in more details in Artigau et al.\ in prep.


The radial velocity (RV) measurements were obtained from the telluric corrected spectra using the line-by-line (LBL) method of \cite{Artigau2022}. In this framework, Doppler shifts are measured on small chunks of the spectrum called `lines' ($\sim$16,000 lines for an M dwarfs observed with SPIRou) using the \cite{Bouchy2001} formalism. With thousands of independent velocity measurements over the full domain, outlying spectral features are readily identified and removed. This is particularly important in the near-infrared where spurious features introduced by complex detector architecture and imperfect telluric correction can significantly bias RV observations \citep{Artigau2022}. The final RV is obtained from the error-weighted average of all valid per-line velocities. The LBL method enabled the mass detection of several TESS planets with SPIRou in recent years (e.g., TOI-1759\,b, \citealt{Martioli2022}; TOI-2136\,b, \citealt{Gan2022}; TOI-1452\,b, \citealt{Cadieux2022}; TOI-1695\,b, \citealt{Kiefer2023}). After subtracting the systemic velocity of about 42.1 km\,s$^{-1}$, we list the LBL RVs of \tar\ in Table~\ref{RVtable}, which have a median uncertainty of 24 m\,s$^{-1}$. As seen in Figure~\ref{RV}, the Doppler signal of the transiting companion is clearly detected. 

\begin{figure}
\includegraphics[width=0.49\textwidth]{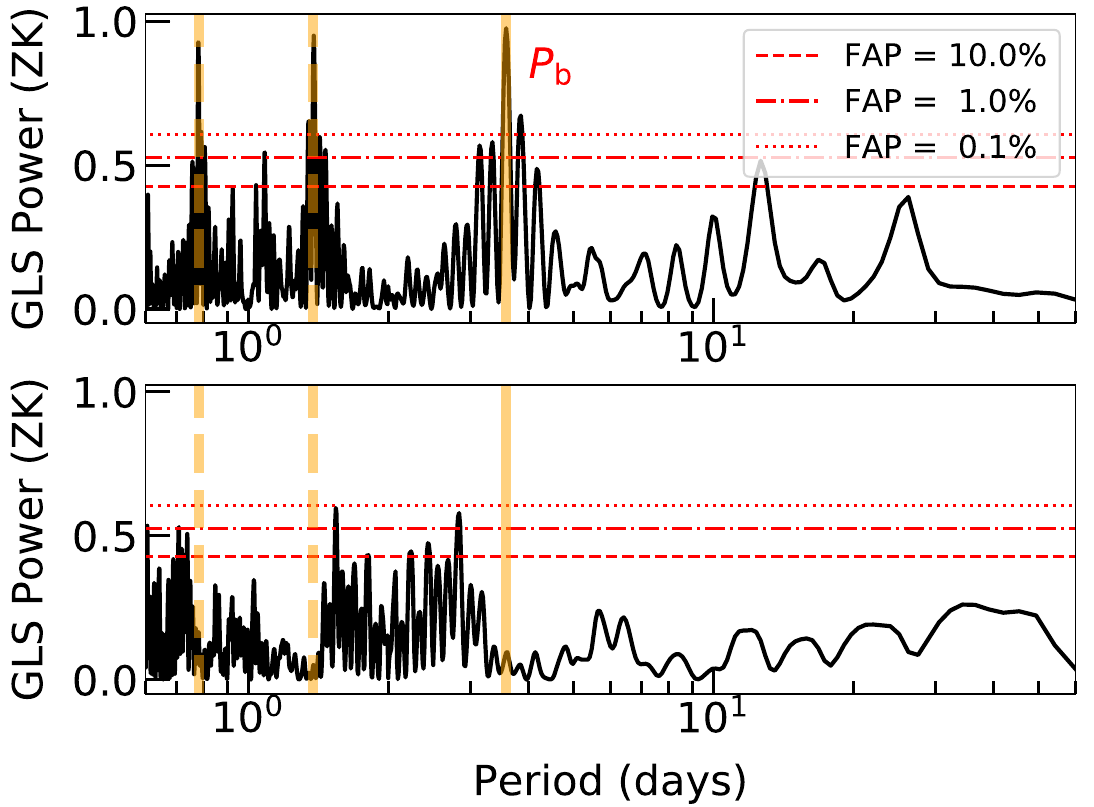}
\caption{{\it Top panel:} The GLS periodogram of SPIRou radial velocities of \tar. {\it Bottom panel:} The GLS periodogram of SPIRou RVs but after subtracting the best-fitting Keplerian model of \tar b. In both panels, the theoretical FAP levels of 10, 1, and 0.1 percent are marked as horizontal dashed, dot–dashed and dotted lines. The orbital period of \tar b is marked as a vertical solid line. Another two significant peaks around 0.78 and 1.38 days (vertical dashed lines) are the signals due to frequency beating. }

\label{RV_gls}
\end{figure}

\begin{figure*}
\includegraphics[width=0.99\textwidth]{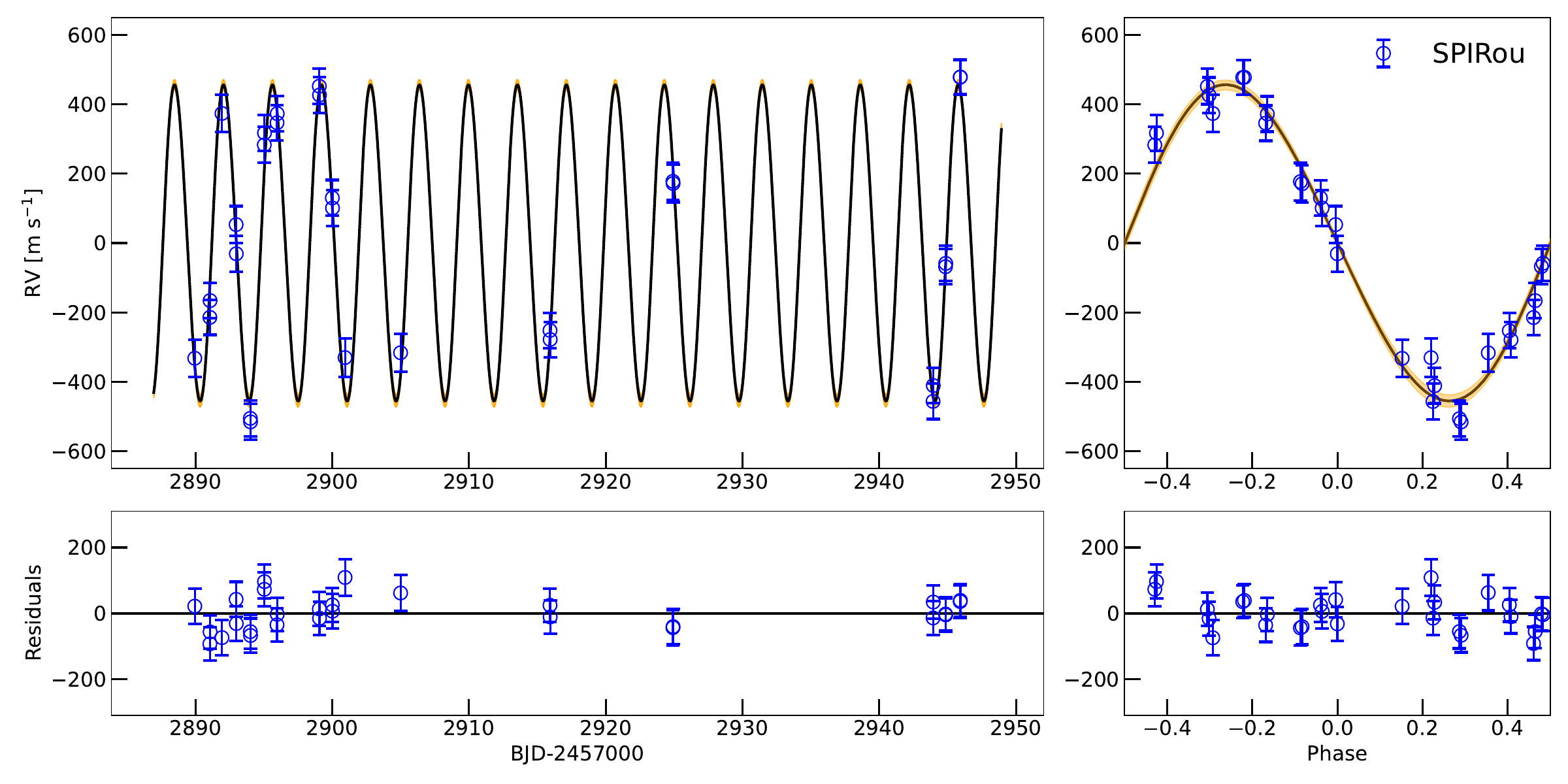}
\caption{{\it Left panel:} The time-series SPIRou radial velocities of \tar. {\it Right panel:} The RVs phased-folded to the orbit period of \tar b. In both panels, the black solid lines are the best Keplerian model from joint-fit. The orange shaded region denotes the $1\sigma$ credible intervals of the RV model. Residuals are plotted below.}
\label{RV}
\end{figure*}

We then performed a generalized Lomb-Scargle (GLS) periodogram \citep{Zechmeister2009} analysis on the SPIRou RVs (see Figure \ref{RV_gls}). A significant peak around 3.58 days is detected at the orbital period of the transiting planet \tar b. In addition, another two signals at 0.78 and 1.38 days can be clearly seen in the periodogram. These two signals come from the alias of the orbital period with the window function, which correspond to the frequency of $f=1+1/3.58$~day$^{-1}$ and $f=1-1/3.58$~day$^{-1}$. After subtracting the best Keplerian model from the joint-fit (see Section \ref{sec:fit}), we find that they disappears and no other significant peaks, which may point to the existence of a secondary planet, show up with FAP below 0.1\%.

\begin{table}[h]
    \caption{SPIRou radial velocities for \tar}
    \begin{tabular}{ccc}
        \hline\hline
        BJD       &RV (m~s$^{-1}$) &$\sigma_{\rm RV}$ (m~s$^{-1}$)      \\\hline
        2459889.94627 &-343.4 &29.7 \\
        2459891.05138 &-225.4 &22.6 \\
        2459891.06557 &-177.0 &22.8 \\
        2459891.93548 &362.6 &28.9 \\
        2459892.96892 &42.1 &28.2 \\
        2459892.98312 &-42.1 &25.4 \\
        2459894.00997 &-516.7 &24.5 \\
        2459894.02416 &-526.7 &25.4 \\
        2459895.02960 &272.1 &25.2 \\
        2459895.04379 &306.3 &25.1 \\
        2459895.96208 &335.8 &23.3 \\
        2459895.97627 &362.0 &23.1 \\
        2459899.05405 &441.1 &24.0 \\
        2459899.06824 &415.7 &24.1 \\
        2459900.00510 &119.1 &24.5 \\
        2459900.01929 &89.8 &26.3 \\
        2459900.93532 &-341.4 &32.9 \\
        2459904.99832 &-327.4 &30.0 \\
        2459915.92131 &-263.0 &23.6 \\
        2459915.93550 &-289.3 &23.2 \\
        2459924.91005 &166.7 &29.1 \\
        2459924.92423 &160.6 &29.3 \\
        2459943.93428 &-467.7 &24.0 \\
        2459943.94847 &-422.0 &23.6 \\
        2459944.84612 &-79.2 &23.0 \\
        2459944.86031 &-69.8 &22.9 \\
        2459945.91587 &467.0 &21.8 \\
        2459945.93007 &467.7 &22.0 \\
         \hline\hline 
    \end{tabular}
    \label{RVtable}
\end{table}

\subsection{High Angular Resolution Imaging}

Close stellar companions (bound or line of sight) can confound exoplanet discoveries in a number of ways.  The detected transit signal might be a false positive due to a background eclipsing binary and even real planet discoveries will yield incorrect stellar and exoplanet parameters if a close companion exists and is unaccounted for \citep{Ciardi-2015ApJ...805...16C,FH1-2017AJ....154...66F, FH2-2020ApJ...898...47F}.
Additionally, the presence of a close companion star leads to the non-detection of small planets residing with the same exoplanetary system \citep{Lester21-2021AJ....162...75L}. Given that nearly one-half of solar-like stars are in binary or multiple star systems, \citep{Matson18-2018AJ....156...31M}  high-resolution imaging provides crucial information toward our understanding of exoplanet formation, dynamics and evolution \citep{Howell2021}. 

\begin{figure}
\includegraphics[width=0.49\textwidth]{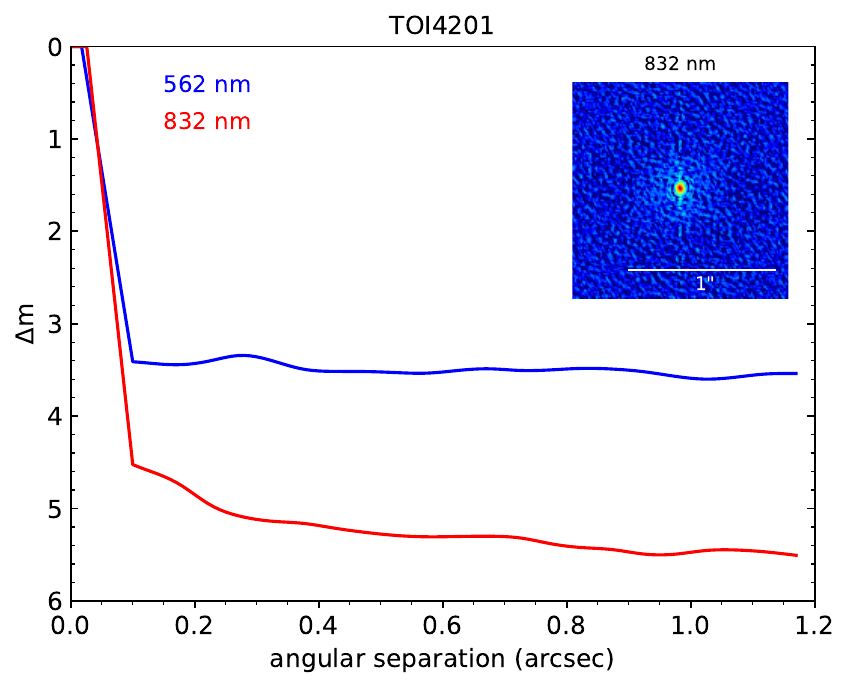}
\caption{Gemini high-resolution speckle imaging 5$\sigma$ contrast limits as a function of angular separation. We show a separate curve for each filter and the 832 nm reconstructed image. TOI-4201 has no close companions within the angular and contrast levels achieved.}
\label{HRI}
\end{figure}

TOI-4201 was observed on 2023 April 25 UT using the Zorro speckle instrument on the Gemini South 8-m telescope \citep{Scott2021, HF-2022FrASS...9.1163H}. Zorro provides simultaneous speckle imaging in two bands (562nm and 832 nm) with output data products including a reconstructed image with robust contrast limits on companion detections. Nine sets of 1000 $\times$ 0.06 second images were obtained and processed in our standard speckle imaging reduction pipeline \citep[see][]{Howell2011}. Figure \ref{HRI} shows our final 5$\sigma$ contrast curves and the 832 nm reconstructed speckle image. 
The observations of TOI-4201 revealed it to have no close companions within the angular and contrast levels achieved.
We find no companion brighter than 4-5 magnitudes below that of the star itself from the 8-m telescope diffraction limit (20 mas) out to 1.2$''$. At the distance of TOI-4201 (d=189 pc), these angular limits correspond to spatial limits of 3.8 to 227 AU.

\section{Stellar Characterization}\label{sec:stellar}

\subsection{Stellar parameters}

We apply three different methods to determine the stellar properties of \tar\ including SED fitting, spectroscopic analysis and empirical relations. We take the weighted stellar radius, mass, effective temperature, and list them in Table~\ref{starparam}.

\subsubsection{SED fitting} 

\begin{figure}
    \centering
    \includegraphics[width=0.49\textwidth]{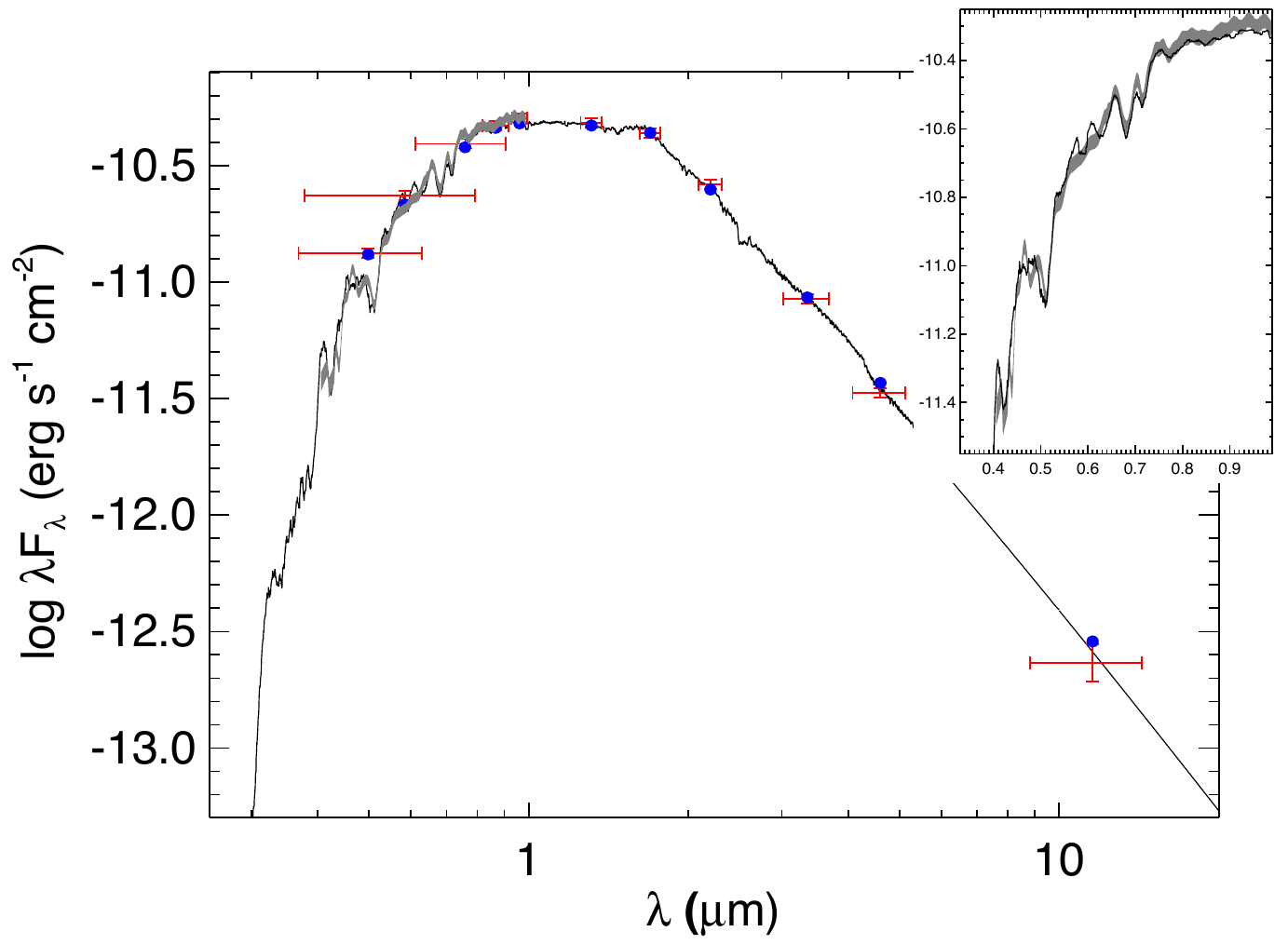}
    \caption{Spectral energy distribution of TOI-4201. Red symbols represent the observed photometric measurements, where the horizontal bars represent the effective width of the passband. Blue symbols are the model fluxes from the best-fit PHOENIX atmosphere model (black). The {\it Gaia\/} spectrum is shown as a grey swathe; the inset focuses on the {\it Gaia\/} spectrum. \label{fig:sed}}
\end{figure}

We carry out a spectral energy distribution (SED) analysis of the star together with the {\it Gaia\/} DR3 parallax \citep[with no systematic offset applied; see, e.g.,][]{StassunTorres:2021}, in order to determine an empirical measurement of the stellar radius, following the procedures described in \citet{Stassun:2016,Stassun:2017,Stassun:2018}. We pulled the $JHK_S$ magnitudes from {\it 2MASS} \citep{Cutri2003}, the W1--W3 magnitudes from {\it WISE} \citep{wright2010}, the $zy$ magnitudes from {\it Pan-STARRS} \citep{Chambers2016}, and the $G$, $G_{\rm BP}, G_{\rm RP}$ magnitudes from {\it Gaia} DR3 \citep{Gaia2022}. Together, the available photometry spans the full stellar SED over the wavelength range 0.4--10~$\mu$m (see Figure~\ref{fig:sed}).  


We perform a fit using PHOENIX stellar atmosphere models \citep{Husser2013}, with the free parameters being the effective temperature ($T_{\rm eff}$) and metallicity ([Fe/H]), as well as the extinction $A_V$, which we limited to maximum line-of-sight value from the Galactic dust maps of \citet{Schlegel:1998}. The resulting fit (Figure~\ref{fig:sed}) has a best-fit $A_V = 0.03 \pm 0.02$, $T_{\rm eff} = 3800 \pm 75$~K, [Fe/H] = $0.2 \pm 0.1$, with a reduced $\chi^2$ of 1.4. The metallicity in particular is further constrained by the {\it Gaia\/} spectrum (see grey swathe in Figure~\ref{fig:sed} and inset). Integrating the (unreddened) model SED gives the bolometric flux at Earth, $F_{\rm bol} = 6.80 \pm 0.16 \times 10^{-11}$ erg~s$^{-1}$~cm$^{-2}$. Taking the $F_{\rm bol}$ and $T_{\rm eff}$ together with the {\it Gaia\/} parallax, gives the stellar radius, $R_\ast = 0.63 \pm 0.03$~R$_\odot$. 



We also independently derive the stellar effective temperature ($T_{\rm eff}$) and radius $R_*$ following the method outlined in \citep{Mann2016}. To briefly summarize, we fit the SED by comparing the available photometry to a grid of late-K and early-M dwarf templates from \citep{Gaidos2014}. To fill in gaps in the templates, we use PHOENIX BT-SETTL models \citep{Allard2013}. The free parameters are the template choice, the scale factor, reddening, model selection ($T_{\rm eff}$ and [M/H]), and three parameters meant to account for systematics if the flux calibration of the templates. We locked the reddening to $E(B-V)<0.1$ based on 3D extinction maps \citep{Green2019}, although this did not change the final fit. We estimate the luminosity from the integral of the absolutely-calibrated spectrum and \gaia\ DR3 parallax, then estimate $R_\ast$ from the Stefan-Boltzmann relation as well as the infrared-flux method \citep{Blackwell1977}. The final resulting fit is $T_{\rm eff}=3784\pm83$\,K and $R_\ast=0.65\pm0.03R_\odot$, consistent with other estimates. 

The planet host sits high on the color-magnitude diagram for a main-sequence M dwarf \citep{Mann2015}, suggesting it is metal-rich ([Fe/H]$>+0.2$ dex). However, deriving a metallicity from this method assumes a single star with no reddening. Instead, we estimate [Fe/H] using the $JHK_S$ photometry as outlined in \cite{Newton2014}. This yielded [Fe/H]$=0.3\pm0.1$, although it may be an underestimate as the host star sits above most of the calibration stars (see Figure 21 in \citealt{Newton2014}).

\subsubsection{Spectroscopic analysis} 

\begin{figure}
    \centering
    \includegraphics[width=1.02\linewidth]{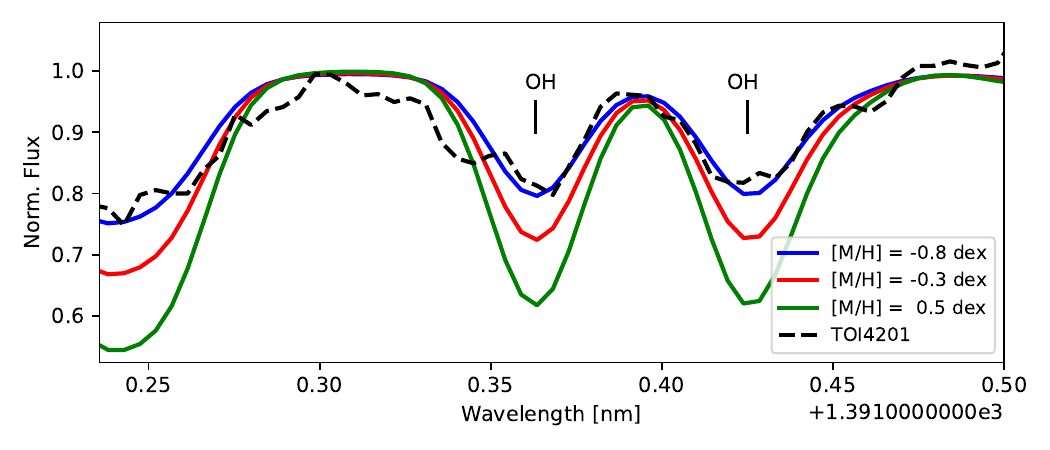}
    \includegraphics[width=0.98\linewidth]{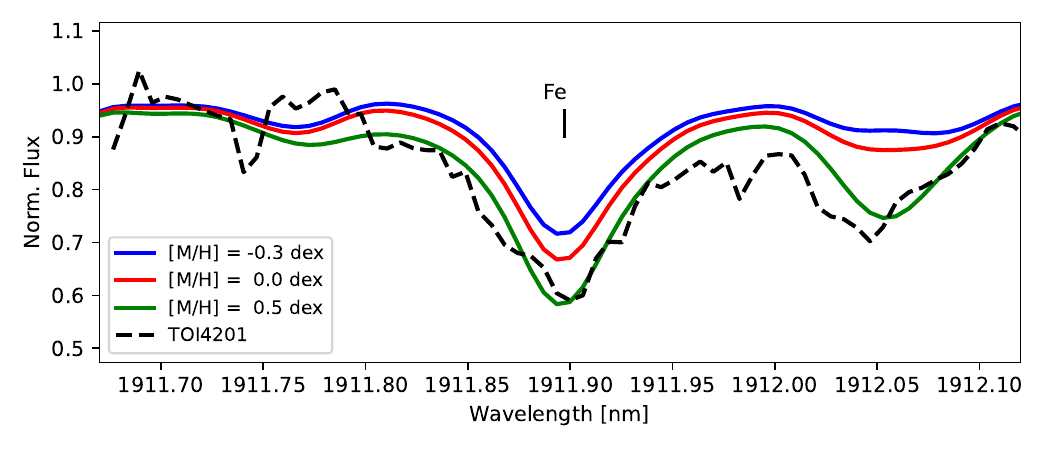}
    \caption{SPIRou observation of OH and Fe I lines of TOI-4201 (black dashed line). The solid lines represent the ACES models for different overall metallicities [M/H] and the fixed $T_{\rm eff}$ of 3800\,K. These plots illustrate the good sensitivity of near-IR high-resolution spectroscopy for constraining the chemical abundances of OH and Fe lines.}
    \label{fig:teff_met}
\end{figure}

The SPIRou high-resolution combined spectrum of \tar\ offers an independent determination of its chemical composition. Such characterization on SPIRou data was also performed on the M4 dwarf TOI-1452 in \cite{Cadieux2022} and additional details on the methodology presented below are given in Jahandar et al.\ in prep. Briefly, the combined spectrum of \tar\ is compared to a 2D grid ($T_{\rm eff}$ and [M/H]) of ACES stellar models (\citealt{Allard2012models}; \citealt{Husser2013}) deconvolved to match SPIRou resolution. We select strong absorption lines matching the models to avoid continuum mismatch and with a known origin, i.e., atomic or molecular and found in the PHOENIX/BT-Settl (\citealt{Allard2012}; \citeyear{Allard2013}) and NIST \citep{Kramida2022} line lists. Fixing the $T_{\rm eff}$ to the value derived above, we performed fits on individual spectral lines to determine the abundances of several chemical species by varying the overall metallicity [M/H]. Since the combined spectrum of \tar\ has a relatively low SNR of $\sim$200 in $H$ band, we also ran Monte Carlo simulations (varying flux values) to assess the robustness of the fits. We find a high stellar iron abundance of $0.52\pm0.08$ dex for \tar\ with 12 Fe I absorption lines, which is consistent with the estimations above from the photometry within $2\sigma$. The final [Fe/H] from the spectroscopic analysis is based on the average of best-fit overall metallicity [M/H] of each Fe line. The uncertainty of [Fe/H] provided in the work is the standard variation of these [M/H] divided by $\sqrt{N-1}$, where N is the total number of lines (12). We note that the error bar might be underestimated since this methodology has an intrinsic RMS of about 0.2 dex \citep[see Figure 11 in][which used the same synthetic model and observed data, although different choice of lines.]{Cristofari2022}. We use the [Fe/H] from SPIRou spectrum analysis for further investigation in the rest of this work and list all metallicity measurements from different sources in Table \ref{starparam}. Meanwhile, we see very shallow OH lines, which indicates a low oxygen abundance (see Figure~\ref{fig:teff_met}). High-resolution spectroscopic observations in optical bands are required to confirm these abundance measurements. 




\subsubsection{Empirical relation}

Based on the $m_{K}$ from 2MASS and parallax from \gaia\ DR3, we obtain the absolute magnitude $M_{K}$ of \tar\ to be $4.98 \pm 0.03$ mag. We then estimate the stellar radius by employing the empirical relation between $R_{\ast}$ and $M_{K}$ derived by \cite{Mann2015}, and we find an $R_\ast$ of $0.61\pm0.02\ R_{\odot}$. This is consistent with the value $0.65\pm0.03\ R_{\odot}$ within $1\sigma$ measured using the relation between stellar angular diameter and color $V-K_s$ described in \cite{Boyajian2014}.

Next, we estimate the stellar effective temperature $T_{\rm eff}$ using two methods. First, we calculate the bolometric correction ${BC}_{K}$ using its relation with $V-J$ from \cite{Mann2015}, and we find ${BC}_{K}=2.55\pm0.04$ mag. Therefore, we measure a bolometric magnitude of $M_{\rm bol}=7.54\pm 0.05$ mag, which results in a bolometric luminosity of $L_{\ast}=0.076\pm0.004\ L_{\odot}$. The effective temperature $3800\pm77$ K is then derived using the Stefan-Boltzmann law. Furthermore, we also obtain an empirical estimation of $T_{\rm eff}=3803
\pm 82$ K using a polynomial relation with stellar colors $V-J$ and $J-H$ \citep{Mann2015}.  

Finally, we obtain the stellar mass $M_{\ast}$ using Equation 2 in \cite{Mann2019} according to the $M_{\ast}$-$M_{K}$ relation \footnote{\url{https://github.com/awmann/M_-M_K-}}. We find $M_{\ast}=0.60\pm0.01\ M_{\odot}$, which agrees with $0.63\pm0.02\ M_{\odot}$ based on the result from \cite{Benedict2016}. We in turn use the mass-radius relation from \cite{Boyajian2012} (see their Eq. 10), and we find a stellar radius of $0.61\pm0.02\ R_{\odot}$, consistent with other results above. 

\begin{table}[h]\scriptsize
    \caption{Summary of stellar parameters for \tar}
    \begin{tabular}{lll}
        \hline\hline
        Parameter       &Value       &Ref. \\\hline
        \it{Main identifiers}                    \\
         TIC                     &$95057860$   &$\rm TIC\ V8^{[1]}$\\
         \gaia\ ID            &$2997312063605005056$ &\gaia\ DR3$^{[2]}$\\
         \it{Equatorial Coordinates} \\
         $\alpha_{\rm J2015.5}$    &06:01:53.93 &$\rm TIC\ V8$\\
         $\delta_{\rm J2015.5}$    &-13:27:40.93   &$\rm TIC\ V8$ \\
         \it{Photometric properties}\\
         $\tess$\ (mag)           &$13.501\pm0.007$   &$\rm TIC\ V8$  \\
         $B$ (mag) &$16.696\pm0.139$ &APASS$^{[3]}$\\
         $V$ (mag) &$15.297\pm0.024$ &APASS\\
         $G$ (mag)           &$14.480\pm0.003$   &\gaia\ DR3   \\
         $G_{\rm BP}$ (mag)           &$15.472\pm0.003$   &\gaia\ DR3   \\
         $G_{\rm RP}$ (mag)           &$13.495\pm0.004$   &\gaia\ DR3   \\
         $J$\ (mag)                    &$12.258\pm0.021$   &2MASS$^{[4]}$\\
         $H$\ (mag)                    &$11.564\pm0.024$   &2MASS \\
         $K$\ (mag)                    &$11.368\pm0.025$    &2MASS \\
         $W$1 (mag)                   &$11.272\pm0.024$   &WISE$^{[5]}$ \\
         $W$2 (mag)                   &$11.301\pm0.021$   &WISE \\
         $W$3 (mag)                   &$11.283\pm0.155$   &WISE \\
         \it{Astrometric properties}\\
         $\varpi$ (mas)              &$5.291\pm0.018$  &\gaia\ DR3  \\
         $\mu_{\alpha}\ ({\rm mas~yr^{-1}})$     &$11.731\pm0.017$   &\gaia\ DR3   \\
         $\mu_{\delta}\ ({\rm mas~yr^{-1}})$     &$6.052\pm0.018$   &\gaia\ DR3  \\
         \it{Stellar parameters} \\
         Spectral Type &$\rm M1.0\pm 0.5$ & This work\\
         RV\ (km~s$^{-1}$)                          &$42.1\pm 0.3$ &This work  \\
         Distance (pc)                &$189.0\pm 0.6$  &This work     \\
         $U_{\rm LSR}$ (km~s$^{-1}$)       &$-23.9\pm 0.2$     &This work\\
         $V_{\rm LSR}$ (km~s$^{-1}$)       &$-17.6\pm 0.2$     &This work\\
         $W_{\rm LSR}$ (km~s$^{-1}$)       &$6.0\pm 0.1$     &This work\\
         $M_{\ast}\ (M_{\odot})$ &$0.61\pm 0.02$ &This work       \\
         $R_{\ast}\ (R_{\odot})$ &$0.63\pm 0.02$ &This work       \\
         $\rho_\ast\ ({\rm g~cm^{-3}})$ &$3.44\pm 0.35$ &This work \\
         $\log g_{\ast}\ ({\rm cgs})$       &$4.64\pm 0.03$  &This work        \\
         $L_{\ast}\ (L_{\odot})$ &$0.076\pm 0.004$  &This work    \\
         $T_{\rm eff}\ ({\rm K})$           &$3794\pm 79$  &This work       \\
         $\rm [Fe/H]$  &$0.30\pm 0.10$ &This work$^{[6]}$ \\
         $\rm [Fe/H]$  &$0.22\pm 0.20$ &This work$^{[7]}$ \\
         $\rm [Fe/H]$  &$0.52\pm 0.08$ &This work$^{[8]}$ \\
         $P_{\rm rot}$\ (days) &$17.3\pm 0.4$ &This work\\
         $\rm Age$ (Gyr) &$0.7-2.0$ &This work\\
         \hline\hline 
    \end{tabular}
    \begin{tablenotes}
    \item[1]  [1]\cite{Stassun2019tic}; [2]\cite{Gaia2022}, [3]\cite{Henden2016}; [4]\cite{Cutri2003}; [5]\cite{wright2010}; 
    [6] From photometric analysis; [7] From spectroscopic analysis of MagE data; [8] From spectroscopic analysis of SPIRou data.
    \end{tablenotes}
    \label{starparam}
\end{table}

\subsection{Stellar rotation and age estimation}

\begin{figure}
\includegraphics[width=0.49\textwidth]{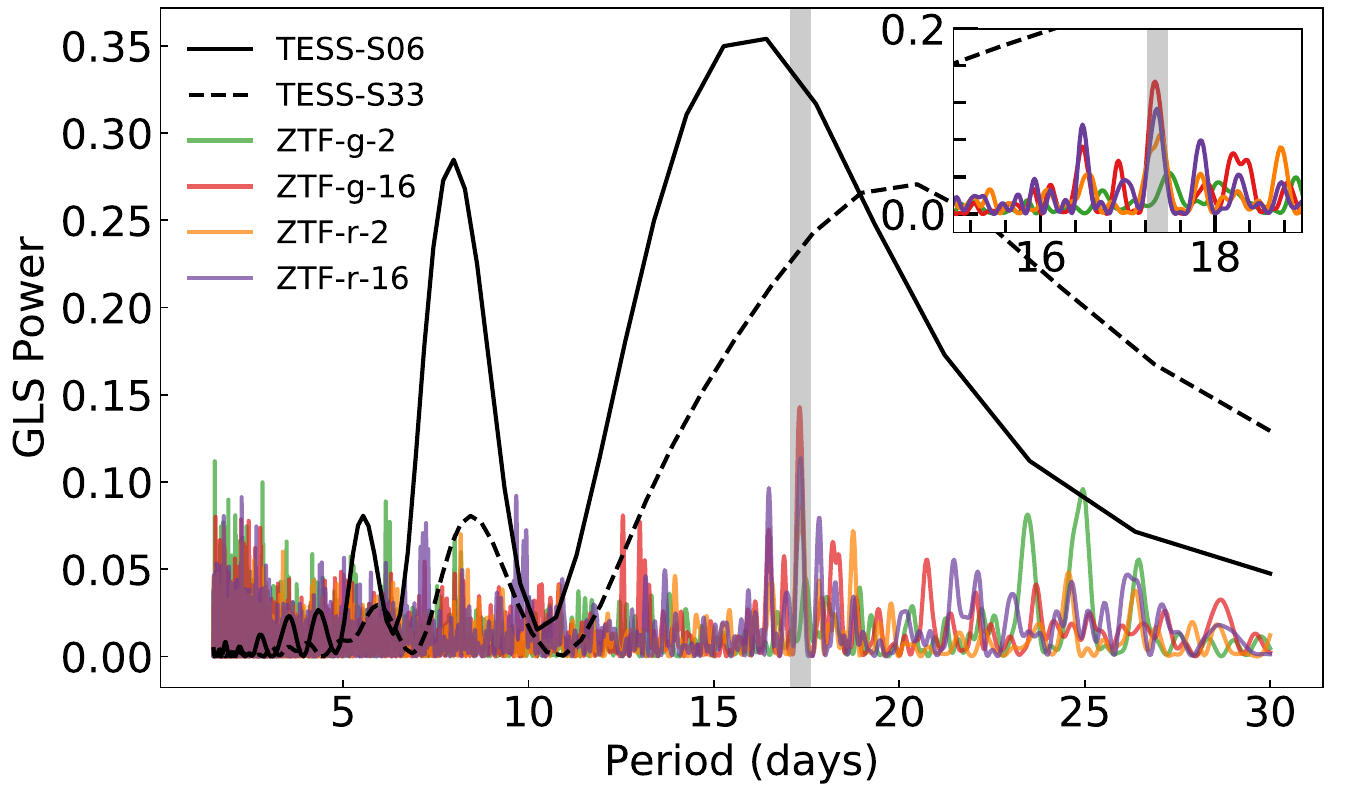}
\caption{The generalized Lomb-Scargle periodograms of the TESS and ZTF photometry. The black solid and dashed lines are the results of TESS while the colored lines are from ZTF, with filter and CCD id shown in the legend. The vertical gray line represents the $\sim 17$ d rotational signal of \tar. A zoomed plot is shown on the top right. }
\label{longterm}
\end{figure}

The TESS light curves of \tar\ that we extracted using a custom aperture from both Sectors 6 and 33 have an obvious baseline variation, which implies a high stellar rotation speed. A similar modulation is also shown in the QLP light curve of Sector 6. We conduct a frequency analysis using the generalized Lomb-Scargle (GLS) periodogram \citep{Zechmeister2009} after removing all in-transit data to measure the stellar rotation period. The result is presented in Figure \ref{longterm}. The light curve from Sector 6 shows a signal around 16.4 days as well as a significant aliasing signal peaking at 8.2 days. While the one from Sector 33 has a broad signal around 20.4 days, similar aliasing signals could also be seen around 8 days. Both the 16.4-day and 20.4-day signals have a false alarm probability (FAP) below 0.1\%. 

Due to the limited length of the TESS observation baseline, the constraint on the rotation period is poor using TESS-only data. Therefore, we further investigate the long-term ground photometry from Zwicky Transient Facility \citep[ZTF;][]{Bellm2019,Masci2019}, and search for periodic modulations. ZTF is mounted on the 48-inch aperture Schmidt-type Telescope at the Palomar Observatory. Using a wide-field camera consisting of 16 CCDs, which results in a field of view of 47 squared deg, ZTF scans the entire Northern sky every two days, providing a large amount of data for time-domain science. \tar\ is observed by ZTF using two different CCDs (id = 2 and 16) in both $g$ and $r$ bands. All of these observations were done with an exposure time of 30s and data are publicly available under DR16. After removing all flux measurements flagged as bad quality, we find that ZTF has (i) 170 measurements from CCD-2 in $g$ band between 2018 Marth 27 and 2022 December 27; (ii) 184 measurements from CCD-16 in $g$ band between 2018 Marth 27 and 2022 December 27; (iii) 206 measurements from CCD-2 in $r$ band between 2018 September 4 and 2022 December 27; and (iv) 183 measurements from CCD-16 in $r$ band between 2018 September 6 and 2022 December 27. We compute the GLS periodogram for each of these datasets, and we show the results in Figure \ref{longterm}. Except for the first one that shows no significant signals with FAP below 10\%, all the other three light curves have a peak around 17.34 days with a FAP of $\sim$0.1\%, $\sim$10\% and $\sim$1\%, which is consistent with the aforementioned findings from TESS data. A forest of other peaks can also be seen between 15 and 20 days, probably due to the poor sampling. 

Based on \kepler\ observations, \cite{McQuillan2014} analyzed the rotation periods of a series of main sequence stars with effective temperatures below 6500 K. They suggested that a typical M star like \tar\ with $T_{\rm eff}$ around 3800 K has a rotation period between 10 and 40 days. In addition, \cite{Newton2016,Newton2018} made use of ground long-term photometric monitoring from the MEarth survey \citep{Irwin2009}, which leads to a conclusion that the rotation periods of early-M stars are generally within 40 days, although their sample has a limited number of early-M stars. Therefore, we attribute this 17.34-day signal to the stellar rotation. 

Building on the stellar rotation period, we next evaluate the stellar age. We first adopt the empirical relation from \cite{Engle2018}, which yields $1.6\pm 0.9$ Gyr, in agreement with the estimation $785\pm85$ Myr from \citep{Mamajek2008}. Additionally, we use the gyrochronology relation reported by \cite{Barnes2007}. Without taking reddening into consideration, we also find an age below 2 Gyr given the stellar color $B-V$ from APASS \citep{Henden2016}. We thus provide a conservative age estimation of \tar\ between 0.7 and 2.0 Gyr. 

\subsection{Stellar kinematic properties}

Following the method described in \cite{Johnson1987}, we calculate the 3-dimensional velocity of \tar\ with respect to the local standard of rest (LSR). We utilize the stellar proper motion ($\mu_\alpha$ and $\mu_\delta$) and parallax ($\varpi$) from \gaia\ DR3 as well as the systemic velocity RV from SPIRou measurement. Taking the solar velocity components relative to the LSR ($U_\odot$, $V_\odot$, $W_\odot$)=(9.58, 10.52, 7.01) km~s$^{-1}$ from the Large Sky Area Multi-Object Fiber Spectroscopic Telescope \citep[LAMOST;][]{Tian2015} into consideration, we obtain a Galactic space motion of ($U_{\rm LSR}$, $V_{\rm LSR}$, $W_{\rm LSR}$)=($-23.9\pm0.2$, $-17.6\pm0.2$, $6.0\pm0.1$) km~s$^{-1}$. We further make use of the \code{BANYAN\ $\Sigma$} algorithm \citep{Gagne2018} to determine the membership probability of \tar\ within young associations. Using the stellar kinematic parameters above, we find that \tar\ is likely a field star and it shows no evidence of cluster membership. We finally follow the procedure in \cite{Bensby2003,Bensby2014}, and estimate a low probability ratio about $P_{\rm thick}/P_{\rm thin}$=1.3\% for \tar\ belonging to the Galactic thick and thin disk, indicating that \tar\ has a thin-disk origin.

\section{Joint-fit}\label{sec:fit}

We carry out a joint-fit of all photometry and SPIRou radial velocities to derive the planetary physical parameters using the \code{juliet} package \citep{juliet}. Basically, we fit light curves using \code{batman} \citep{Kreidberg2015} while the RV model is generated with \code{radvel} \citep{Fulton2018}. We obtain the posteriors of all parameters through a dynamic nested sampling with \code{dynesty} \citep{Speagle2019}. 

We set Gaussian normal priors on both the orbital period and mid-transit epoch, centering at the values we found in our transit search, with a $1\sigma$ value of 0.1 days. Following the parametrization for the planet-to-star radius ratio $p$ and impact parameter $b$ in \cite{Espinoza2018}, we efficiently sample their physical parameter spaces by fitting $r_{1}$ and $r_{2}$ instead and allow them uniformly vary between 0 and 1. Since we use a relatively large box aperture to extract the \tess\ photometry, light contamination from other stars especially the nearby star \gaia\ 2997312063605005952 located within the aperture with $G=16.8$~mag ($\Delta G=2.3$~mag) should be considered (see Figure~\ref{FOV_aper_TESS}). Therefore, we fit a dilution factor\footnote{The dilution factor is defined as 1/(1+$A_{D}$), where $A_D$ is the light contamination ratio. We use the estimate from TICV8 and set a Gaussian prior.} for the \tess\ photometry while we fix it at 1 for ground-based observations as nearby stars around \tar\ in ground images are deblended. We opt to adopt a quadratic limb darkening law for the \tess\ photometry and a linear law for ground data \citep{Kipping2013}. Finally, we include a jitter term for every photometric dataset to account for additional white noise. 

In terms of the RV modeling, we fit a standard Keplerian orbit, leaving $e\sin \omega$ and $e\cos \omega$ as free parameters. We place wide informative uniform priors on the RV semi-amplitude $K$, the systematic velocity $\mu_{\rm SPIRou}$ as well as the RV jitter $\sigma_{\rm SPIRou}$. Due to the short time span, we do not consider linear and quadratic RV trends here, both of which are fixed at 0. The joint fit reveals that the planet has a radius of $1.22\pm0.04\ R_J$ with a mass of $2.48\pm0.09\ M_J$ on a nearly circular orbit. We list the prior settings and the posteriors of all parameters in Table~\ref{allpriors}. We show the phase-folded data along with the best-fit transit and RV models in Figures~\ref{ground} and \ref{RV}.

\begin{table*}\scriptsize
    {\renewcommand{\arraystretch}{1.05}
    \caption{Parameter priors and the best-fit values along with the 68\% credibility intervals in the final joint fit for \tar. $\mathcal{N}$($\mu\ ,\ \sigma^{2}$) means a normal prior with mean $\mu$ and standard deviation $\sigma$. $\mathcal{U}$(a\ , \ b) stands for a uniform prior between $a$ and $b$. $\mathcal{LU}$(a\ , \ b) is a log-uniform prior between $a$ and $b$. $\mathcal{TN}$($\mu$\ ,\ $\sigma^{2}$, \ a, \ b) represents a truncated normal prior ranging from $a$ to $b$.}
    \begin{tabular}{lccr}
        \hline\hline
        Parameter       &Prior &Value    &Description\\\hline
        \it{Orbit parameters}\\
        $P$ (days)  &$\mathcal{N}$ ($3.6$\ ,\ $0.1$)  &$3.5819194\pm0.0000011$
        &Orbital period.\\
        $T_{c}$ (BJD-2457000) &$\mathcal{N}$ ($1470.9716$\ ,\ $0.1$) &$1470.9619\pm0.0004$ &Mid-Transit time.\\
        $r_1$ &$\mathcal{U}$ ($0$\ ,\ $1$) &$0.6662\pm0.0109$ &Parametrization for p and b.\\
        $r_2$ &$\mathcal{U}$ ($0$\ ,\ $1$) &$0.1955\pm0.0008$ &Parametrization for p and b.\\
        $e\sin \omega$ &$\mathcal{U}$ ($-1$\ ,\ $1$) &$-0.0402\pm0.0155$ &Parametrization for $e$ and $\omega$.\\
        $e\cos \omega$ &$\mathcal{U}$ ($-1$\ ,\ $1$) &$0.0006\pm0.0106$ &Parametrization for $e$ and $\omega$.\\
        $K$ (m~s$^{-1}$) &$\mathcal{U}$ ($0$\ ,\ $600$) &$456.3\pm12.5$ &RV semi-amplitude.\\
        $\mu_{\rm SPIRou}$ (m~s$^{-1}$) &$\mathcal{U}$ ($-500$\ ,\ $500$) &$-11.2\pm 8.8$ &Systemic velocity.\\
        $\sigma_{\rm SPIRou}$ (m~s$^{-1}$) &$\mathcal{U}$ ($0$\ ,\ $100$) &$45.2\pm 7.1$ &RV jitter.\\\hline
        \it{Stellar parameter}\\
        $\rho_{\ast}$ (kg~m$^{-3}$) &$\mathcal{LU}$ ($10^{2}$\ ,\ $10^{5}$) &$3786\pm146$ &Stellar Density.\\\hline
        \it{Dilution factors}\\
        $D_{\rm TESS\ S06}$ &$\mathcal{TN}$ ($0.9$\ ,\ $0.1^{2}$,\ 0\ ,\ 1) &$0.85\pm0.01$ & \\
        $D_{\rm TESS\ S33}$ &$\mathcal{TN}$ ($0.9$\ ,\ $0.1^{2}$,\ 0\ ,\ 1) &$0.85\pm0.01$ & \\
        $D_{\rm ground}$ &1 (Fixed) &$\cdots$ & \\\hline
        \it{Limb-darkening coefficients}\\
        $q_{\rm 1,TESS\ S06}$ &$\mathcal{U}$ ($0$\ ,\ $1$) &$0.16\pm0.06$ & \\
        $q_{\rm 2,TESS\ S06}$ &$\mathcal{U}$ ($0$\ ,\ $1$) &$0.65\pm0.23$ & \\
        $q_{\rm 1,TESS\ S33}$ &$\mathcal{U}$ ($0$\ ,\ $1$) &$0.32\pm0.12$ & \\
        $q_{\rm 2,TESS\ S33}$ &$\mathcal{U}$ ($0$\ ,\ $1$) &$0.19\pm0.12$ & \\
        $q_{\rm LCO,SAAO,g}$ &$\mathcal{U}$ ($0$\ ,\ $1$) &$0.55\pm0.27$ & \\
        $q_{\rm LCO,SAAO,i}$ &$\mathcal{U}$ ($0$\ ,\ $1$) &$0.19\pm0.12$ & \\
        $q_{\rm LCO,CTIO,g}$ &$\mathcal{U}$ ($0$\ ,\ $1$) &$0.75\pm0.09$ & \\
        $q_{\rm LCO,CTIO,i}$ &$\mathcal{U}$ ($0$\ ,\ $1$) &$0.30\pm0.08$ & \\
        $q_{\rm MuSCAT,g}$ &$\mathcal{U}$ ($0$\ ,\ $1$) &$0.54\pm0.06$ & \\
        $q_{\rm MuSCAT,r}$ &$\mathcal{U}$ ($0$\ ,\ $1$) &$0.62\pm0.03$ & \\
        $q_{\rm MuSCAT,z}$ &$\mathcal{U}$ ($0$\ ,\ $1$) &$0.31\pm0.04$ & \\
        $q_{\rm SPECULOOS,z}$ &$\mathcal{U}$ ($0$\ ,\ $1$) &$0.29\pm0.04$ & \\\hline
        \it{Photometric jitter}\\
        $\sigma_{\rm TESS\ S06}$ (ppm) &$\mathcal{LU}$ ($10^{-6}$\ ,\ $10^{5}$)  &$8.3^{+54.2}_{-8.3}$ & \\
        $\sigma_{\rm TESS\ S33}$ (ppm) &$\mathcal{LU}$ ($10^{-6}$\ ,\ $10^{5}$)  &$0.1^{+32.7}_{-0.1}$ & \\
        $\sigma_{\rm LCO,SAAO,g}$ (ppm) &$\mathcal{LU}$ ($10^{-6}$\ ,\ $10^{5}$)  &$15.7^{+43.6}_{-14.9}$ & \\
        $\sigma_{\rm LCO,SAAO,i}$ (ppm) &$\mathcal{LU}$ ($10^{-6}$\ ,\ $10^{5}$)  &$31.8^{+99.5}_{-30.9}$ & \\
        $\sigma_{\rm LCO,CTIO,g}$ (ppm) &$\mathcal{LU}$ ($10^{-6}$\ ,\ $10^{5}$)  &$2663.7^{+599.2}_{-476.9}$ & \\
        $\sigma_{\rm LCO,CTIO,i}$ (ppm) &$\mathcal{LU}$ ($10^{-6}$\ ,\ $10^{5}$)  &$2000.5^{+374.3}_{-310.7}$ & \\
        $\sigma_{\rm MuSCAT,g}$ (ppm) &$\mathcal{LU}$ ($10^{-6}$\ ,\ $10^{5}$)  &$12.4^{+19.6}_{-11.8}$ & \\
        $\sigma_{\rm MuSCAT,r}$ (ppm) &$\mathcal{LU}$ ($10^{-6}$\ ,\ $10^{5}$)  &$38.8^{+48.3}_{-37.6}$ & \\
        $\sigma_{\rm MuSCAT,z}$ (ppm) &$\mathcal{LU}$ ($10^{-6}$\ ,\ $10^{5}$)  &$2000.5^{+19.4}_{-12.5}$ & \\
        $\sigma_{\rm SPECULOOS,z}$ (ppm) &$\mathcal{LU}$ ($10^{-6}$\ ,\ $10^{5}$)  &$2309.0^{+172.8}_{-169.4}$ & \\\hline
        \it{Derived parameters}\\
        $R_{p}/R_{\ast}$ &$\cdots$ &$0.1955\pm0.0008$ &Scaled planet radius.\\
        $a/R_{\ast}$ &$\cdots$ &$13.694\pm0.151$ &Scaled semi-major axis.\\
        $a$ (AU) &$\cdots$ &$0.040\pm 0.001$ &Semi-major axis.\\
        $b$ &$\cdots$ &$0.499\pm0.016$ &Impact parameter.\\
        $i$ (degrees) &$\cdots$ &$88.0\pm0.1$ &Orbital inclination.\\
        $e$ &$\cdots$ &$0.041\pm0.015$ &Orbital eccentricity.\\
        $\omega$ (degrees) &$\cdots$ &$130.0\pm44.5$ &Argument of periapsis.\\\hline
        \it{Planetary physical parameters}\\
        $R_{p}$ ($R_{J}$) &$\cdots$ &$1.22\pm 0.04$ &Planet radius.\\
        $M_{p}$ ($M_{J}$) &$\cdots$ &$2.48\pm 0.09$ &Planet Mass.\\
        $\rho_{p}$ (g~cm$^{-3}$) &$\cdots$ &$1.82\pm 0.19$ &Planet density.\\
        $T_{\rm eq}^{[1]}$ (K) &$\cdots$ &$725\pm 20$ &Equilibrium temperature.\\
        $S$ ($S_{\oplus}$) &$\cdots$ &$45.2\pm 3.2$ &Planetary Insolation.\\
        \hline\hline
    \label{allpriors}    
    \end{tabular}}
    \begin{tablenotes}
       \item[1]  [1]\ We do not consider heat distribution between the dayside and nightside here and assume albedo $A_B=0$. 
    \end{tablenotes}
\end{table*}


\section{Discussions}\label{sec:dis}

\subsection{\tar b: A massive and dense giant planet}

\begin{figure}
\centering
\includegraphics[width=1\linewidth]{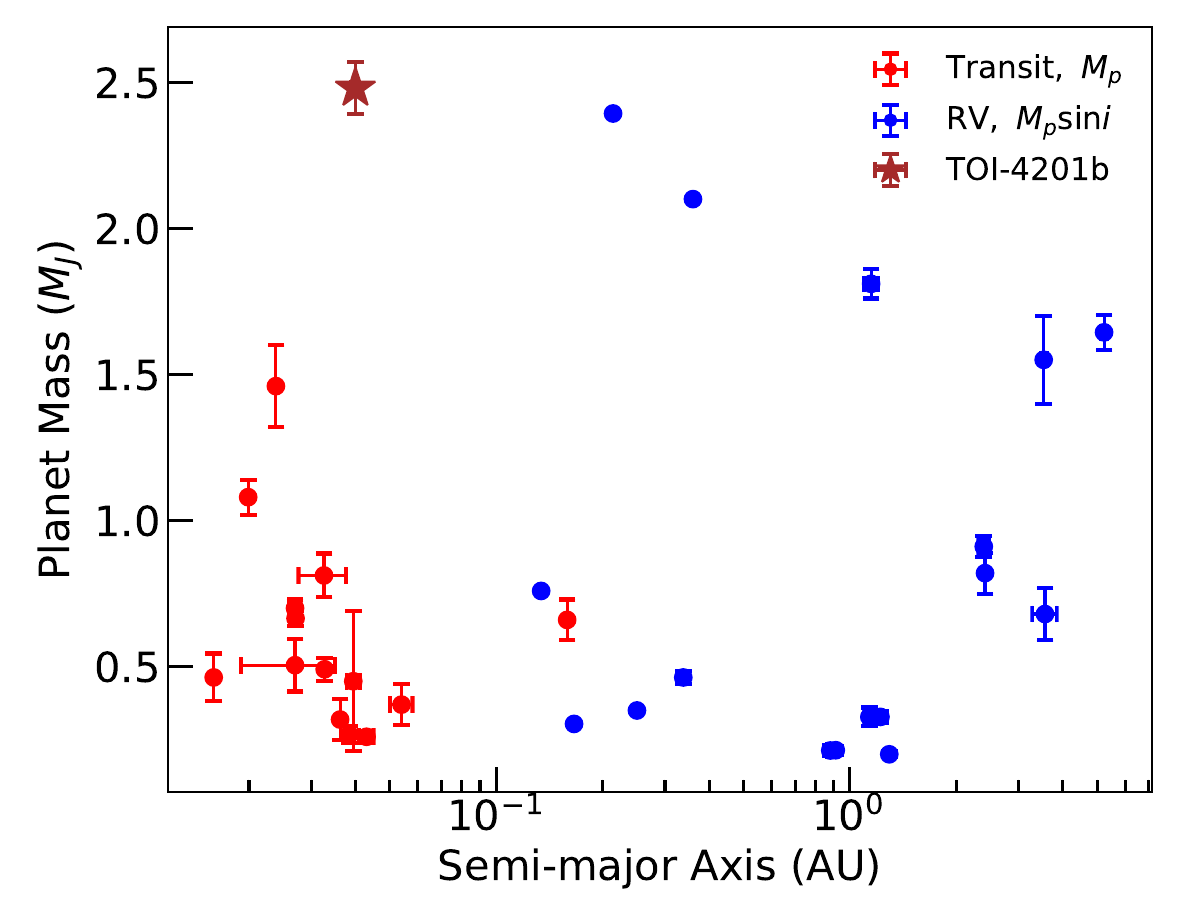}
\includegraphics[width=1\linewidth]{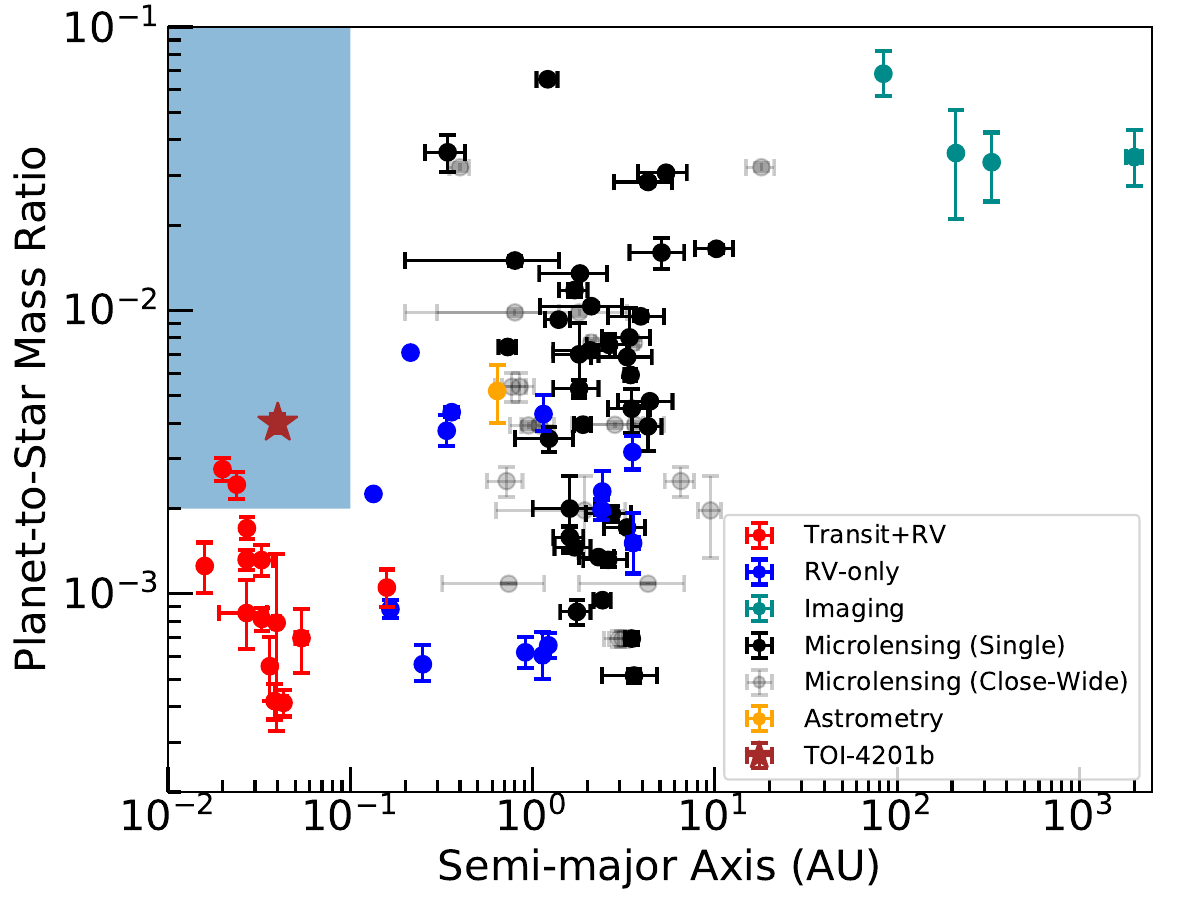}
\caption{{\it Top panel:} The mass and semi-major axis diagram of all confirmed gas giants around M dwarfs from transit (red) and RV (blue) surveys. {\it Bottom panel:} The planet-to-star mass ratio and semi-major axis distribution of gas giants around M dwarfs colored by the detection techniques. The blue shaded region is the mass ratio paucity region proposed by \cite{Gantoi530}. In both panels, \tar b is marked as a brown star.}
\label{amq}
\end{figure}

With a mass of around $2.5\ M_J$, \tar b is the most massive hot Jupiter around an M dwarf known so far (see the top panel of Figure~\ref{amq}). It stands out from other similar systems (M dwarf + hot Jupiter) with a much lower companion mass around $0.5\ M_J$. Since most host stars of those systems are also early-M dwarfs, the reason why \tar b is more massive than others is unclear. In addition, \tar b also has the highest planet-to-star mass ratio of about $4\times10^{-3}$ among M dwarf hot Jupiter systems (see the bottom panel of Figure~\ref{amq}). Compared with warm and cold Jupiters detected by RV and microlensing surveys, it turns out that such massive close-in Jupiters are lacking around M dwarfs. Together with HATS-74Ab \citep{Jordan2022} and TOI-5205b \citep{Kanodia2023}, \tar b is the third planet situated in the mass ratio paucity region ($q\geq 2\times 10^{-3}$ and $a\leq 0.1$ AU), an unpopulated area proposed by \cite{Gantoi530}, making it an interesting object for theories of planet formation. 

\begin{figure}
\includegraphics[width=0.49\textwidth]{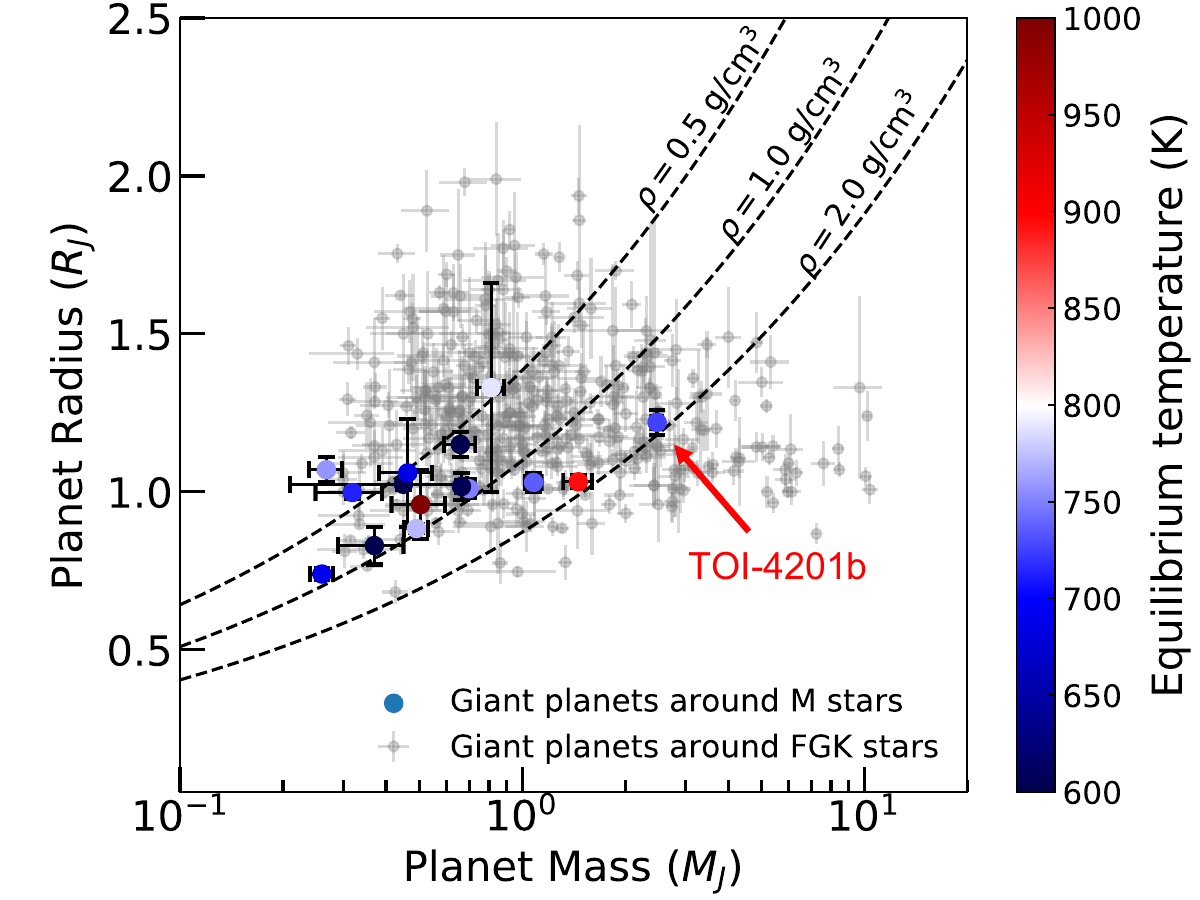}
\caption{Planet mass and radius diagram. The gray dots are giant planets around FGK stars. The colored dots are all transiting gas giants around M dwarfs with mass measurements. Different colors represent different equilibrium temperatures. Three constant density curves ($\rho$=0.5, 1.0, 2.0 g~cm$^{-3}$) are shown as dashed lines for reference. With a density around 1.8 g~cm$^{-3}$, \tar b is one of the densest gas giants transiting M dwarfs.}
\label{MR}
\end{figure}

Figure~\ref{MR} shows the Mass-Radius diagram of all confirmed transiting gas giants around FGKM dwarfs. Unlike those giant planets around FGK stars that have both diverse radius and mass, gas giants around M dwarfs have masses spanning orders of magnitude but with relatively concentrated radius, which is mainly due to the low incident flux they received. Because these giant planets around M dwarfs are not strongly irradiated by the host star, they are good targets to study interior structures.

\subsection{Interior structure of \tar b}

The mean density of \tar b is about $1.82\pm0.19$\,g~cm$^{-3}$, significantly higher than other observed gas giants around M-dwarfs, as is shown in Figure~\ref{MR}. The high density of the planet can be entirely attributed to hydrogen compression in the massive planet, and not of high metal content, as in the case of lower mass planets, since high $Z$ would make the planet radius too small. Interior-evolution modeling suggests metal content between 0-25 $M_{\oplus}$ in the interior of \tar b, depending on the metal distribution in the interior, its history, and its energy state. 

According to the standard core accretion models \citep[e.g.,][]{fortney07,baraffe08}, the mean density of \tar b is consistent with pure H,\ He composition and lighter than that, i.e. inflated radius. 
We perform dedicated structure-evolution models of \tar b under various conditions, based on \cite{vazan13,vazan15} in order to fit its radius-mass-age parameters. 
Since atmospheric properties of \tar b are not available, we consider albedo values of 0-0.3, and atmospheric opacity metallicity between solar and ten times solar \citep{freedman14}. 
Evolution models covered a wide range of parameters of cold and hot start conditions \citep[e.g.,][]{marley07}, and various metal distributions in the interior. The models suggest a core mass of 0-5 Earth masses and a metal-poor envelope. Alternative scenarios that involve post-formation giant impact allow larger metal content. In such cases \tar b can contain up to 25 Earth masses of metals, gradually distributed in the deep interior, surrounded by a metal-poor envelope. 



\begin{figure}
\includegraphics[width=0.49\textwidth]{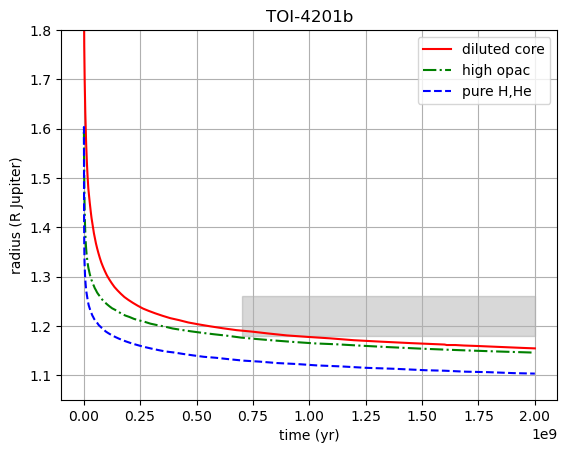}
\caption{Radius evolution of three \tar b like planets. The observed values of radius-age appear in grey box. The evolution models are of a metal-free planet (dashed blue), hot diluted core of 25$M_{\oplus}$ (red), and of 5$M_{\oplus}$ core with high atmospheric opacity (green). Models emphasize the requirement of additional energy sources to explain the observed properties of \tar b.}
\label{fig:Rev}
\end{figure}

In Figure~\ref{fig:Rev}, we show the radius evolution of three \tar b mass planets. Under standard evolution conditions \tar b is an inflated hot Jupiter, since its observed radius-age (grey box) is larger than any model with efficient (adiabatic) cooling, even if the model is of a metal-free planet (dashed blue).  
The hot diluted core model (red) simulates post energetic giant impact scenario that eroded parts of the core, injecting energy and slowing the cooling of the planet. Yet, the maximum metal mass in this scenario cannot exceed 25$M_{\oplus}$, and the model meets the observed values only at a limited range. A classical core-envelope model with 5$M_{\oplus}$ core and high atmospheric opacity of ten times solar (green) barely fits the observed properties. These models indicate a difficulty to fit observations with high accuracy without additional energy sources. 

Based on the power-law relation from \cite{Thorngren2016}, we compute the total heavy element mass required for \tar b, and we obtain a high value around $100\ M_\oplus$ under the assumption that the planet is not affected by inflation mechanisms. Apart from a typical $10\ M_\oplus$ core occupied in \cite{Thorngren2016}, the rest of the metals with a mass of about $90\ M_{\oplus}$ are supposed to go directly into the gas envelope. However, the resulting low metal content (up to 3\%) of TOI-4201b from our interior structure modeling makes this planet challenging in terms of planet formation theory. In this system, despite the high metallicity of \tar, the giant planet appears to be a metal-poor planet and its bulk metallicity is at most identical to its star metallicity. This finding challenges the classical core-accretion trend of the stellar-planet metallicity \citep{Thorngren2016}. The formation path of \tar b - a massive metal-poor planet around a metal-rich low-mass star - remains unclear. Given the low metallicity of the planet, planet formation by gravitational instability cannot be ruled out \citep{boss11}. 

\subsection{Potential heating mechanisms}

The discrepancy between the host star iron abundance and planet metallicity ($Z_{\rm planet}< 0.03$ and $Z_{\rm star}=0.067\pm0.012$) could be slightly mitigated if additional heating processes have an effect. Such processes may inject energy and slow down the cooling, leading to an inflated planet radius. Although the low equilibrium temperature ($<1000$\,K) makes Ohmic dissipation less probable \citep{Batygin2010}, we propose three possible pathways of planet heating. 

\subsubsection{Tidal heating}

The short distance from the star suggests that tidal heating may introduce non-negligible energy into the planet's interior, as it went through the circularization of its orbit \citep{Leconte2010}. The low orbital eccentricity of $0.041\pm0.015$ also suggests that the tidal effect has an influence on the planet. We calculate the tidal circularization timescale of \tar b using the equation from \cite{Goldreich1966}, which yields a $\tau_{e}=0.68$ Gyr, assuming a typical planet quality factor $Q_{p}\sim10^{6}$ of hot Jupiters. This is very close to the lower bound age estimation $0.7-2.0$ Gyr for the host star. If this is the case, it favors the younger age limit for the system and the tidal heating probably took place until recently.

\subsubsection{Gas giant merger}

Another possible scenario is a head-on collision of two less massive gas giants \citep{Li2010,Liu2015}, which will result in a coalescence of solid cores and gaseous envelopes without substantial material loss. Due to the high energy generated during the impact, the final merger remnant might be puffed up. This scenario also accounts for the high mass of the planet. 


Planet collision should be more likely around metal-rich stars. With high solid density protoplanetary disks, they are likely to have multiple giant planets at the beginning. Such a configuration has also been found around M dwarfs, for example GJ 317, a metal-rich ([Fe/H]=$0.30\pm0.08$ dex) M star hosting two cold Jupiters \citep{Anglada2012GJ317,Feng2020}. Violent planet–planet interactions will probably then occur through gravitational perturbations \citep{Rasio1996}. Under certain conditions, one of these formed giant planets will be scattered inward to large eccentricities \citep{Dawson2013}. While in some extreme cases, the instability may cause close encounters if planets have sufficiently small spacings, which will result in a direct collision \citep{Li2021}. 

As a violent process, one would naturally expect to see a misalignment between the spin axis of the host star and the orbital angular momentum axis of the planet, induced by planet-planet scattering \citep{Liu2020}. Observations have shown that several hot Jupiters have high obliquities (projected spin-orbit angle), which may be left during the interactions between planets \citep{Winn2015}. If \tar b did undergo a dynamically hot formation history, we may be able to find clues through studying the stellar obliquity \citep{Stefansson2022}. 


\subsubsection{Embryo capture by a gas giant}



An alternative route to inject energy is through embryo capture. At first, there probably exists a proto gas giant outside the snow line accompanied with several rocky embryos such as super Earths in the inner disk. During the inward migration of the cold giant planet \citep{Lin1996}, these interior embryos would be trapped by the mean motion resonance (MMR) of the gas giant, leading to an orbital decay along the path. The planetary resonant perturbation probably then excites the orbits of these unstable embryos \citep{Zhou2007,SWang2021}. If their eccentricity damping rate is fast, we will see a short-period giant accompanied by close-in super Earths. However, the orbits of rocky embryos may crisscross that of the migrating gas giant if the damping timescale is long. They will pass through the envelope of the proto gas giant and collides into its center core. Similar to the aforementioned pathway, the energy produced by the impact may inflate the gas giant.


\begin{figure}
    \includegraphics[width=1\linewidth]{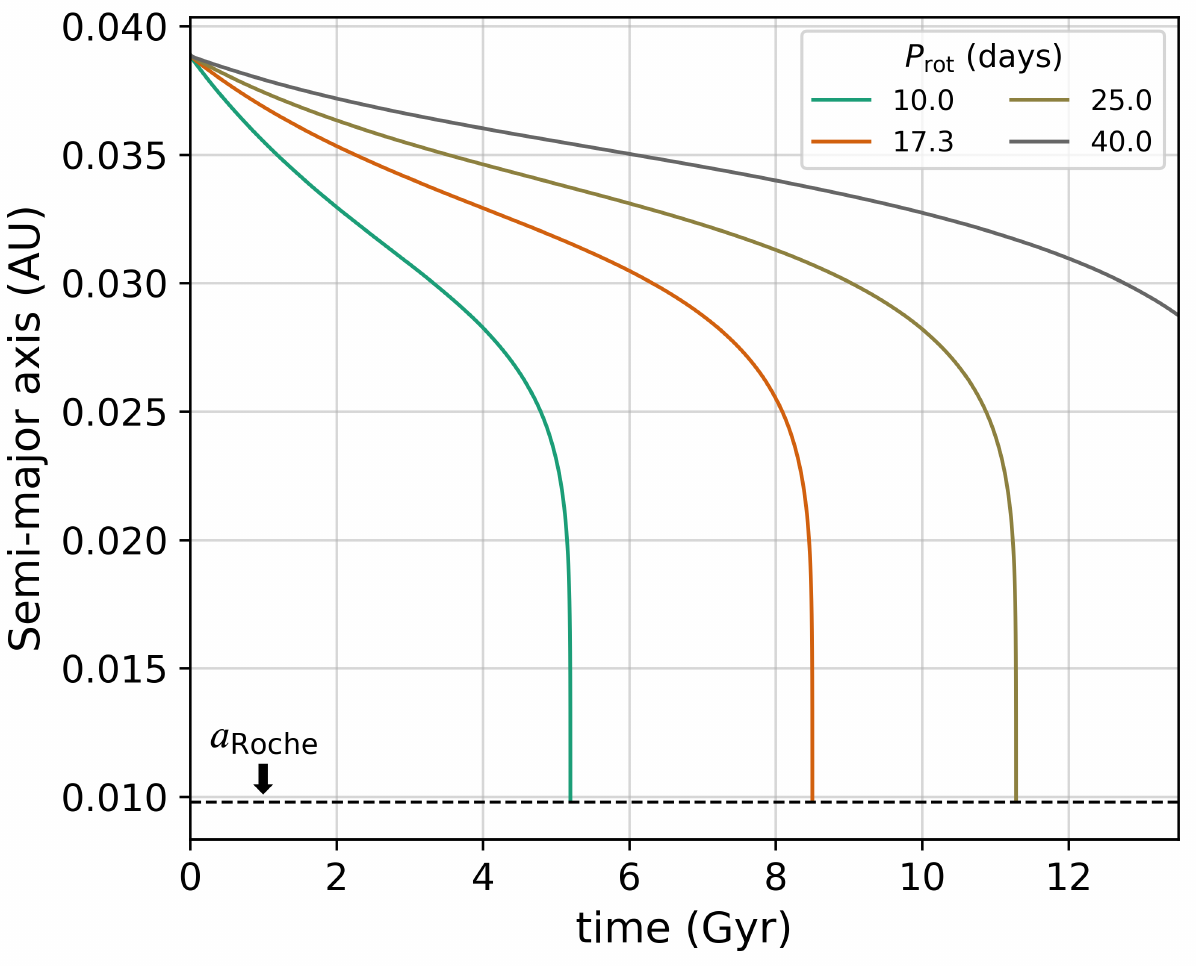}
    \includegraphics[width=1\linewidth]{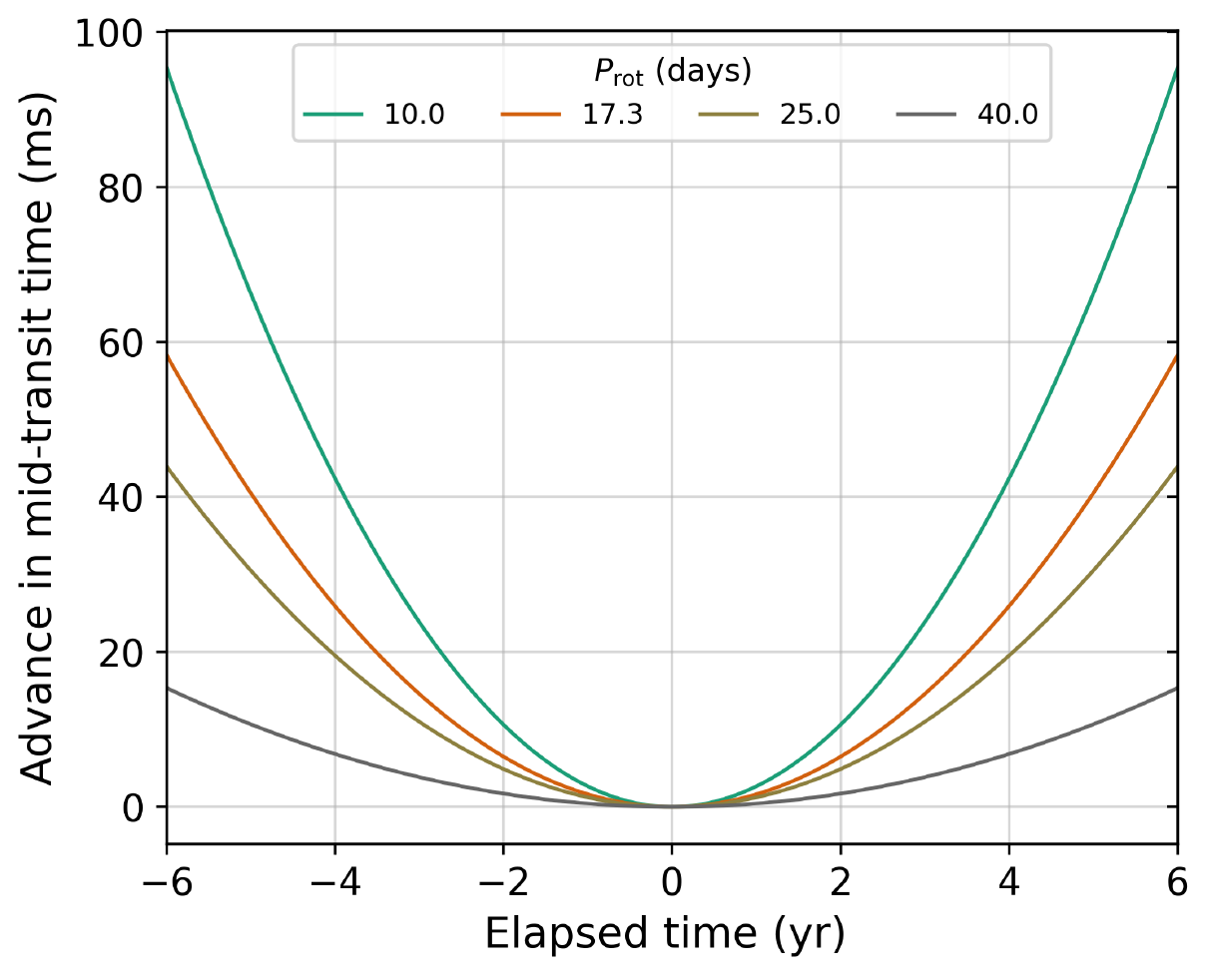}
    \caption{{\it Top panel:} Semi-major axis as a function of time. {\it Bottom panel:} advance in mid-transit for \tar b for an observation baseline of six years measured from today. For both panels, the colors represent different stellar rotation periods ($P_{\rm rot}$) from 10 to 40 days, as expected for M dwarfs \citep{McQuillan2014}. The orange line stands for the measured $P_{\rm rot}$ in this work.}
    \label{fig:tidaldecay}
\end{figure}

A planetary system with both outer gas giants and inner small planets is not unusual. \cite{Huang2016} found that nearly half of warm Jupiters co-exist with low-mass planets. In addition, recent work from \cite{DWu2023} also reported that at least $12\pm6$\% hot Jupiters as well as $\geq$ $70\pm16$\% warm Jupiters have nearby planetary companions by investigating the transit timing variation signals, which probably support this hypothesis. Although the number is still limited, such a planetary configuration has also been found around M dwarfs through RV surveys \citep[e.g., GJ 876;][]{Rivera2010}. Since the disk migration and embryo capture are quiescent, unlike the previously mentioned process, here we expect a well-aligned system with a low stellar obliquity. Such gentle gas disk migrations are also unlikely to produce large eccentricities of giant planets \citep{Dunhill2013}, like what is expected for \tar b.

\subsection{Tidally induced evolution}
As \tar b is a new addition to the growing number of giant planets orbiting M dwarfs (see \citealt{Alvarado-Montes2022}), studying its tidal evolution could allow us to shed some light on the energy dissipation of M-dwarf hosts and put some constraints on stellar and planetary interiors. We study the tidal evolution of \tar b under the formalism used by \cite{Alvarado-Montes2022} to understand the orbital decay of giant planets orbiting M dwarfs due to the dissipation of inertial waves (IWs) in convective envelopes and internal gravity waves (IGWs) in stellar radiative regions. For this particular system, IWs were not excited in the convective envelope due to the short orbital period of the planet and the slow rotation of the host star (i.e. $P\ngeq\frac{P_\mathrm{rot}}{2}$; see \citealt{Barker2020}).

IGWs in the stellar radiative regions are not expected to be damped due to wave breaking, as a planetary mass threshold (i.e. critical mass) $M_{\rm crit}>10^3\,M_J$ would be necessary for such a mechanism to occur for the given stellar age and stellar mass (see $M_{\rm crit}$ values in Figure 9 in \citealt{Barker2020}). However, if IGWs are somehow excited (e.g., via radiative diffusion), we predict a stellar quality factor $Q'=Q'_{\rm IGW}\approx2.6\times10^7$. This could also be the case in later stages of stellar evolution as $M_{\rm crit}$ becomes smaller, thus producing fully damped IGWs due to wave breaking. For this tidal quality factor, we found that \tar b would undergo orbital decay from its initial position to the Roche limit in a timescale of $\sim$ 8 Gyr for a 17.3-day stellar rotation period. The orbital decay for other stellar rotation periods is also shown in Figure \ref{fig:tidaldecay} (top panel), and the migration timescales resemble those of other short-period giant planets undergoing orbital decay for $P_\mathrm{rot}<40$ days (see e.g., \citealt{Brown2011,Alvarado2019}).

Using the results from the tidal simulations performed for \tar b (top panel in Figure \ref{fig:tidaldecay}), we computed the rate of change of the orbital period of \tar b as $\dot{P}=0.03185^{+0.00049}_{-0.00044}$ ms yr$^{-1}$. Using these results, we calculated the advance in the mid-transit time of \tar b (bottom panel in Figure \ref{fig:tidaldecay}) where it can be seen that an advance of $\sim$ 60 ms is expected for the next six years of observations, being three orders of magnitude smaller than those calculated for ultra-short-period Jupiters (see e.g., \citealt{McCormac2019, Alvarado2021}). It is worth noting that we studied the tidal effects for \tar b assuming a co-planar system, so future observations (see next subsection) can help us refine our calculations by using a formalism where the evolution of the projected obliquity $\lambda$ would also be included.

\subsection{Prospects for future observations}

Based on the mass correlation from \cite{Thorngren2016}, \tar b is supposed to contain a large amount of metal in the envelope but we would expect to see a smaller planet radius in that case. Our interior structure model, instead, suggests the envelope is metal-poor. It motivates future atmospheric characterization using space telescopes like JWST to study the chemical composition in the planet atmosphere. Therefore, we compute the Transmission Spectroscopy Metric \citep[TSM;][]{Kempton2018}, which is a quality factor of an object suitable for atmospheric composition studies. We find a $\rm TSM=24\pm6$, much smaller than the threshold of 90 for high-quality Jovian-like planets recommended by \cite{Kempton2018}, indicating that \tar b is a challenging case but still possible with multi-visit observations. 

In addition, studying stellar obliquity may help us trace back the planet dynamic history. The host star shows a clear rotation signal around 17.3 days, corresponding to a stellar rotation speed of $v\sin i\sim 1.8$ km~s$^{-1}$. To probe the opportunity of detecting the Rossiter–McLaughlin (RM) effect \citep{Rossiter1924,McLaughlin1924} and measure the projected spin orbit angle, we estimate the RM signal amplitude to a first order using 
\begin{equation}
    A_{\rm RM} \sim \frac{2}{3}(R_{p}/R_{\ast})^{2}\sqrt{1-b^{2}}\times v\sin i,
\end{equation}
where $b$ is the impact parameter. The moderate impact parameter also eliminates the usual covariance between $v\sin i$ and the projected obliquity $\lambda$. Coupled with results from the joint fit, we find an $A_{\rm RM}$ about 44 m/s. Although the host star is faint and the transit duration is short, the RM observation is still possible with ground spectroscopic facilities on large telescopes like MAROON-X and ESPRESSO. 



\subsection{Metallicity preference of hot and warm Jupiters around M dwarfs}

\begin{figure}
\centering
\includegraphics[width=0.49\textwidth]{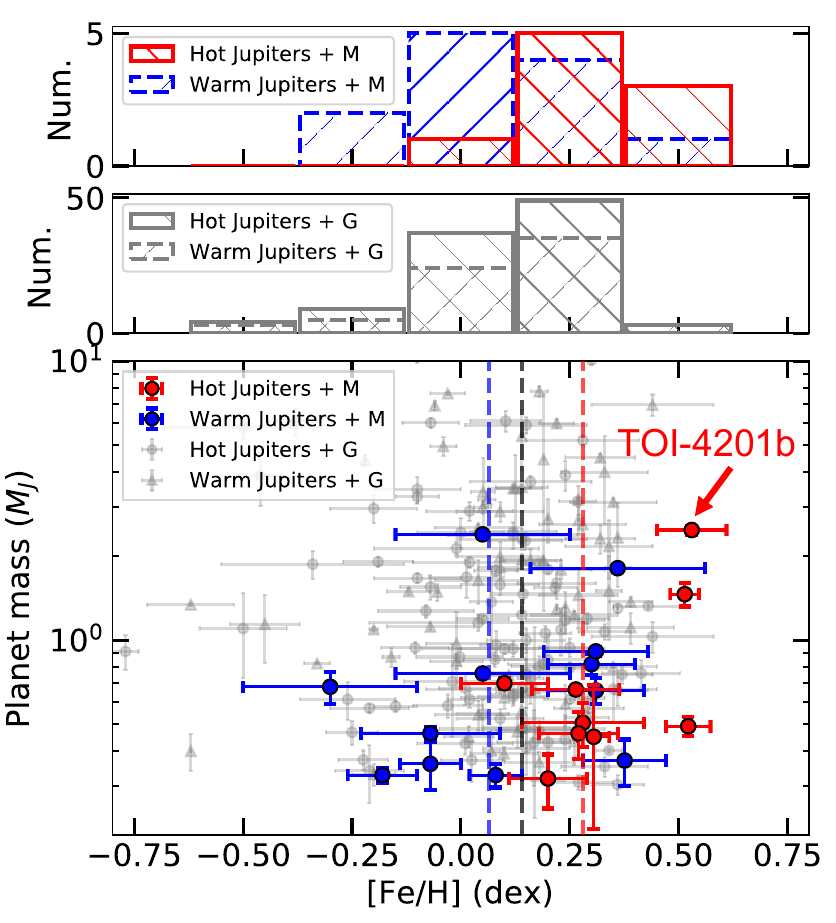}
\caption{Planet mass ($M_p$ or $M_p \sin i$) versus host star metallicity. The red and blue dots are hot ($a/R_{\ast}< 20$) and warm ($a/R_{\ast}\geq 20$) Jupiters around M dwarfs. The vertical red and blue dashed lines are the median metallicity of two samples: 0.28 and 0.06 dex. The background gray dots and triangles are hot and warm Jupiters around G-type stars under the same definition. Both of them have a median iron abundance of 0.14 dex, shown as a vertical black line. The top two panels are the histogram of these four samples.}
\label{iron}
\end{figure}

In order to compare the planet-metallicity dependence to probe the similarity of the planet formation history, we next investigate the metallicity distribution of four planet groups: HJG, WJG, HJM and WJM, corresponding to hot and warm Jupiters around G ($0.90 \leq M_{\ast}\leq 1.06\ M_{\odot}$) and M dwarfs based on the discoveries retrieved from the NASA Exoplanet Archive \citep{Akeson2013}. To take the planet insolation into consideration, here we define the hot and warm Jupiter groups with a scaled semi-major axis ($a/R_{\ast}$) and planet mass boundary cut. We designate hot Jupiters as planets having mass above 0.3 $M_{J}$ and $a/R_{\ast}<20$ while cold Jupiters as those with $a/R_{\ast}>20$. Under this definition, our sample contains 103 HJG, 70 WJG, 9 HJM and 12 WJM. First, both hot and warm Jupiters are likely to form around metal-rich G-type stars. These two samples have a similar preference for stellar metallicity, both of which have a median [Fe/H] around 0.14 dex as shown in Figure~\ref{iron}. Interestingly, most hot Jupiters are orbiting M dwarfs with even higher super-solar metallicity ($\sim$0.28 dex). Such a strong dependence between giant planets and stellar metallicity favors the core accretion formation mechanism \citep{Santos2004,Fischer2005,Sousa2011,WWang2018}. Meanwhile, the iron abundances of M dwarfs that host warm Jupiters have a much wider distribution, from $-0.3$ to 0.3 dex with a median of 0.06 dex. Therefore, the formation of warm Jupiters around M dwarfs is probably less sensitive to stellar metallicity, indicating a different formation pathway that is less dependent on metallicity. But it is still possible that this feature is biased due to the limited M dwarf sample size or different facilities and methods used to measure the iron abundance. Future observations and homogeneous spectroscopic analysis may help draw a firm conclusion. 

\section{Conclusions}\label{sec:con}

In this paper, we report the discovery and characterization of \tar b, a massive and dense hot Jupiter transiting an early-M dwarf. \tar b was first alerted as a planet candidate based on the TESS data, and the planetary nature was then confirmed through ground-based photometric, spectroscopic and imaging observations. \tar b has a radius of $1.22\pm  0.04\ R_J$ with a mass of $2.48\pm0.09\ M_J$. It orbits the host M star every 3.58 days on a nearly circular orbit ($e=0.041\pm0.015$). The bulk density $\rho_{p}=1.82\pm0.19$\,g~cm$^{-3}$ of \tar b makes it one of the most massive and densest hot Jupiters around M dwarfs. Although the host star has super-solar metallicity, interior structure modeling suggests that the \tar b is metal-poor, which differs from the classical picture of positive stellar-planet metallicity correlation. Planet formation scenarios more typical for the core accretion model can be envisioned if additional energy sources, like tidal heating or giant impacts, operate to inflate the planet radius. Future studies on the planet atmosphere and stellar obliquity may shed light on its formation and evolution.


We also compare the stellar metallicity distributions of four planet groups: hot and warm Jupiters around G and M stars. We find that the M dwarfs hosting hot Jupiters have a higher mean metallicity than G dwarfs, favoring the core accretion scenario. Warm Jupiters around M dwarfs, instead, show a weak dependence on stellar iron abundance, which perhaps indicates a different formation story. 

\section{Acknowledgments}
We thank Beibei Liu and Haochang Jiang for the useful discussions on planet formation and scattering. We are grateful to Coel Hellier for the insights regarding the WASP data. 

This work is partly supported by the National Science Foundation of China (Grant No. 12133005).
This research uses data obtained through the Telescope Access Program (TAP), which has been funded by the TAP member institutes.
The authors acknowledge the Tsinghua Astrophysics High-Performance Computing platform at Tsinghua University for providing computational and data storage resources that have contributed to the research results reported within this paper.

Based on observations obtained at the Canada-France-Hawaii Telescope (CFHT) which is operated from the summit of Maunakea by the National Research Council of Canada, the Institut National des Sciences de l'Univers of the Centre National de la Recherche Scientifique of France, and the University of Hawaii. The observations at the Canada-France-Hawaii Telescope were performed with care and respect from the summit of Maunakea which is a significant cultural and historic site. Based on observations obtained with SPIRou, an international project led by Institut de Recherche en Astrophysique et Planétologie, Toulouse, France.

This work is partly supported by the Natural Science and Engineering Research Council of Canada and the Institute for Research on Exoplanets through the Trottier Family Foundation. This work makes use of observations from the Las Cumbres Observatory global telescope network and it is partly supported by JSPS KAKENHI Grant Numbers JP17H04574, JP18H05439, and JST CREST Grant Number JPMJCR1761.

Some of the observations in this paper made use of the High-Resolution Imaging instrument Zorro and were obtained under Gemini LLP Proposal Number: GN/S-2021A-LP-105. Zorro was funded by the NASA Exoplanet Exploration Program and built at the NASA Ames Research Center by Steve B. Howell, Nic Scott, Elliott P. Horch, and Emmett Quigley. Zorro was mounted on the Gemini South telescope of the international Gemini Observatory, a program of NSF’s OIR Lab, which is managed by the Association of Universities for Research in Astronomy (AURA) under a cooperative agreement with the National Science Foundation. on behalf of the Gemini partnership: the National Science Foundation (United States), National Research Council (Canada), Agencia Nacional de Investigación y Desarrollo (Chile), Ministerio de Ciencia, Tecnología e Innovación (Argentina), Ministério da Ciência, Tecnologia, Inovações e Comunicações (Brazil), and Korea Astronomy and Space Science Institute (Republic of Korea).

Based on observations obtained with the Samuel Oschin 48-inch Telescope at the Palomar Observatory as part of the Zwicky
Transient Facility project. ZTF is supported by the National Science Foundation under Grant No. AST-1440341 and a
collaboration including Caltech, IPAC, the Weizmann Institute for Science, the Oskar Klein Center at Stockholm University, the
University of Maryland, the University of Washington, Deutsches Elektronen-Synchrotron and Humboldt University, Los Alamos
National Laboratories, the TANGO Consortium of Taiwan, the University of Wisconsin at Milwaukee, and Lawrence Berkeley
National Laboratories. Operations are conducted by COO, IPAC, and UW.

This work makes use of observations from the LCOGT network. Part of the LCOGT telescope time was granted by NOIRLab through the Mid-Scale Innovations Program (MSIP). MSIP is funded by NSF.

This paper includes data gathered with the 6.5 meter Magellan Telescopes located at Las Campanas Observatory, Chile.

The ULiege's contribution to SPECULOOS has received funding from the European Research Council under the European Union's Seventh Framework Programme (FP/2007-2013) (grant Agreement n$^\circ$ 336480/SPECULOOS), from the Balzan Prize and Francqui Foundations, from the Belgian Scientific Research Foundation (F.R.S.-FNRS; grant n$^\circ$ T.0109.20), from the University of Liege, and from the ARC grant for Concerted Research Actions financed by the Wallonia-Brussels Federation. SPECULOOS-North has received financial support from the Heising-Simons Foundation and from Dr. and Mrs. Colin Masson and Dr. Peter A. Gilman. 

The postdoctoral fellowship of KB is funded by F.R.S.-FNRS grant T.0109.20 and by the Francqui Foundation.
This publication benefits from the support of the French Community of Belgium in the context of the FRIA Doctoral Grant awarded to MT. J.d.W. and MIT gratefully acknowledge financial support from the Heising-Simons Foundation, Dr. and Mrs. Colin Masson and Dr. Peter A. Gilman for Artemis, the first telescope of the SPECULOOS network situated in Tenerife, Spain. BVR thanks the Heising-Simons Foundation for support. MG is FNRS-F.R.S Research Director. KAC and SQ acknowledges support from the TESS mission via subaward s3449 from MIT. SQ acknowledges support from the TESS GI Program under award 80NSSC21K1056.

Funding for the TESS mission is provided by NASA's Science Mission Directorate. This research has made use of the Exoplanet Follow-up Observation Program website, which is operated by the California Institute of Technology, under contract with the National Aeronautics and Space Administration under the Exoplanet Exploration Program. This paper includes data collected by the \tess\ mission, which are publicly available from the Mikulski Archive for Space Telescopes\ (MAST). 
This work has made use of data from the European Space Agency (ESA) mission
{\it Gaia} (\url{https://www.cosmos.esa.int/gaia}), processed by the {\it Gaia} Data Processing and Analysis Consortium (DPAC,
\url{https://www.cosmos.esa.int/web/gaia/dpac/consortium}). Funding for the DPAC has been provided by national institutions, in particular the institutions participating in the {\it Gaia} Multilateral Agreement.
This work made use of \texttt{tpfplotter} by J. Lillo-Box (publicly available in www.github.com/jlillo/tpfplotter), which also made use of the python packages \texttt{astropy}, \texttt{lightkurve}, \texttt{matplotlib} and \texttt{numpy}.


\facilities{TESS, Gaia, CFHT/SPIRou, Magellan/MagE, Gemini/Zorro, LCOGT, MuSCAT, SPECULOOS, ZTF}

\software{lightkurve \citep{lightkurvecollaboration}, astropy \citep{2013A&A...558A..33A,2018AJ....156..123A}, AstroImageJ \citep{Collins2017}, juliet \citep{juliet}, batman \citep{Kreidberg2015}, radvel \citep{Fulton2018}, tpfplotter \citep{Aller2020}}

\bibliography{planet}{}
\bibliographystyle{aasjournal}



\end{document}